\documentclass[%
 reprint,
superscriptaddress,
amsmath,amssymb,
aps,
prb,
floatfix
]{revtex4-2}

\usepackage{multirow}
\usepackage{graphicx}
\usepackage{dcolumn}
\usepackage{bm}
\usepackage[version=3]{mhchem}
\usepackage{nicefrac}

\usepackage{color}
\usepackage{dcolumn}
\usepackage{amssymb}
\usepackage{url}
\usepackage[colorlinks=True,linkcolor=red,citecolor=blue,urlcolor=blue]{hyperref}
\usepackage{xfrac}
\usepackage{tabularx}
\usepackage{xspace}
\usepackage{amsmath}
\usepackage[normalem]{ulem}
\usepackage{mathtools}

\def\be{\begin{equation}}
\def\ee{\end{equation}}
\def\bea{\begin{eqnarray}}
\def\eea{\end{eqnarray}}
\def\ba{\begin{array}}
\def\ea{\end{array}}

\def\bar1{\overline{1}}

\begin{document}
\title{Effective spin-1 breathing kagome Hamiltonian induced by the exchange hierarchy in the maple leaf mineral bluebellite}
\author{Pratyay Ghosh}
\affiliation{Institut f\"ur Theoretische Physik und Astrophysik and W\"urzburg-Dresden Cluster of Excellence ct.qmat, Universit\"at W\"urzburg,
Am Hubland Campus S\"ud, W\"urzburg 97074, Germany}

\author{Tobias M\"uller}
\affiliation{Institut f\"ur Theoretische Physik und Astrophysik and W\"urzburg-Dresden Cluster of Excellence ct.qmat, Julius-Maximilians-Universit\"at W\"urzburg, Am Hubland Campus S\"ud, W\"urzburg 97074, Germany}

\author{Yasir Iqbal}
\affiliation{Department of Physics and Quantum Centers in Diamond and Emerging Materials (QuCenDiEM) group, Indian Institute of Technology Madras, Chennai 600036, India}

\author{Ronny Thomale}
\affiliation{Institut f\"ur Theoretische Physik und Astrophysik and W\"urzburg-Dresden Cluster of Excellence ct.qmat, Julius-Maximilians-Universit\"at W\"urzburg, Am Hubland Campus S\"ud, W\"urzburg 97074, Germany}
\affiliation{Department of Physics and Quantum Centers in Diamond and Emerging Materials (QuCenDiEM) group, Indian Institute of Technology Madras, Chennai 600036, India}

\author{Harald O. Jeschke}
\affiliation{Research Institute for Interdisciplinary Science, Okayama University, Okayama 700-8530, Japan}
\affiliation{Department of Physics and Quantum Centers in Diamond and Emerging Materials (QuCenDiEM) group, Indian Institute of Technology Madras, Chennai 600036, India}

\begin{abstract}
  As a highly frustrated model Hamiltonian with an exact dimer ground state, the Heisenberg antiferromagnet on the maple leaf lattice is of high theoretical interest, and a material realization is intensely sought after. We determine the magnetic Hamiltonian of the copper mineral bluebellite using density functional theory based energy mapping. As a consequence of the significant distortion of the spin $S=1/2$ maple leaf lattice, we find two of the five distinct nearest neighbor couplings to be ferromagnetic. Solution of this Hamiltonian with density matrix renormalization group calculations points us to the surprising insight that this particular imperfect maple leaf lattice, due to the strongly ferromagnetic Cu$^{2+}$ dimer, realizes an effective $S=1$ breathing kagome Hamiltonian. In fact, this is another highly interesting Hamiltonian which has rarely been realized in materials. Analysis of the effective model within a bond-operator formalism allows us to identify a valence bond solid ground state and to extract thermodynamic quantities using a low-energy bosonic mean-field theory. We resolve the puzzle of the apparent one-dimensional character of bluebellite as our calculated specific heat has a Bonner-Fisher-like shape, in good agreement with experiment.
  
\end{abstract}

\maketitle

Triangular motifs in quantum antiferromagnets are a source of geometric frustration and lead to highly nontrivial emergent phenomena like quantum spin liquids~\cite{Balents2010}. Starting with the triangular lattice, site depletion leads to new lattices which have lower coordination number but potentially more frustration~\cite{Richter2004,Lu2018}. For example, the kagome lattice is obtained by a $1/4$ site-depletion of the triangular lattice and has coordination number of four; it also hosts some of the most intensively studied spin liquid candidates~\cite{Mendels2011}. In a somewhat more exotic manner, the maple leaf lattice can be viewed as a one-seventh site-depleted triangular lattice and has a coordination number of five~\cite{Betts1995}. The uniform nearest-neighbor Heisenberg antiferromagnetic model on this lattice has been solved via exact diagonalization and other techniques~\cite{Schmalfuss2002,Farnell2018} and found a magnetically ordered ground state. Recently, in an analytic work on the model with a bond anisotropy, Ghosh \emph{et. al.} established an exact dimer ground state~\cite{Ghosh2022}, making the maple leaf lattice the only other two-dimensional lattice with uniform tiling that admits an exact dimer ground state besides the widely known Shastry-Sutherland model~\cite{Shastry1981} which has an extremely rich phenomenology and phase diagram~\cite{Jimenez2021,Shi2022,deconf-SSM,Bound-state-SSM}. While \ce{SrCu2(BO3)2} has been found to be an extremely good representation of the Shastry-Sutherland Hamiltonian, a material realizing the model proposed by Ghosh \emph{et. al.} on the maple leaf lattice has yet to be identified. Candidates involving quantum spins are the copper minerals~\cite{Norman2018,Inosov2018} spangolite \ce{Cu6Al(SO4)(OH)12Cl.3H2O}~\cite{Hawthorne1993}, sabelliite \ce{Cu2ZnAsO4(OH)3}~\cite{Olmi1995},  mojaveite \ce{Cu6TeO4(OH)9Cl}~\cite{Mills2014}, fuetterite \ce{Pb3Cu6TeO6(OH)7Cl5}~\cite{Kampf2013} and finally bluebellite \ce{Cu6IO3(OH)10Cl}~\cite{Mills2014}. Magnetic properties have been characterized experimentally for spangolite~\cite{Fennell2011} and bluebellite~\cite{Haraguchi2021}, but the magnetic Hamiltonians for any of these maple leaf compounds remains to be established.

\begin{figure*}
    \includegraphics[width=\textwidth]{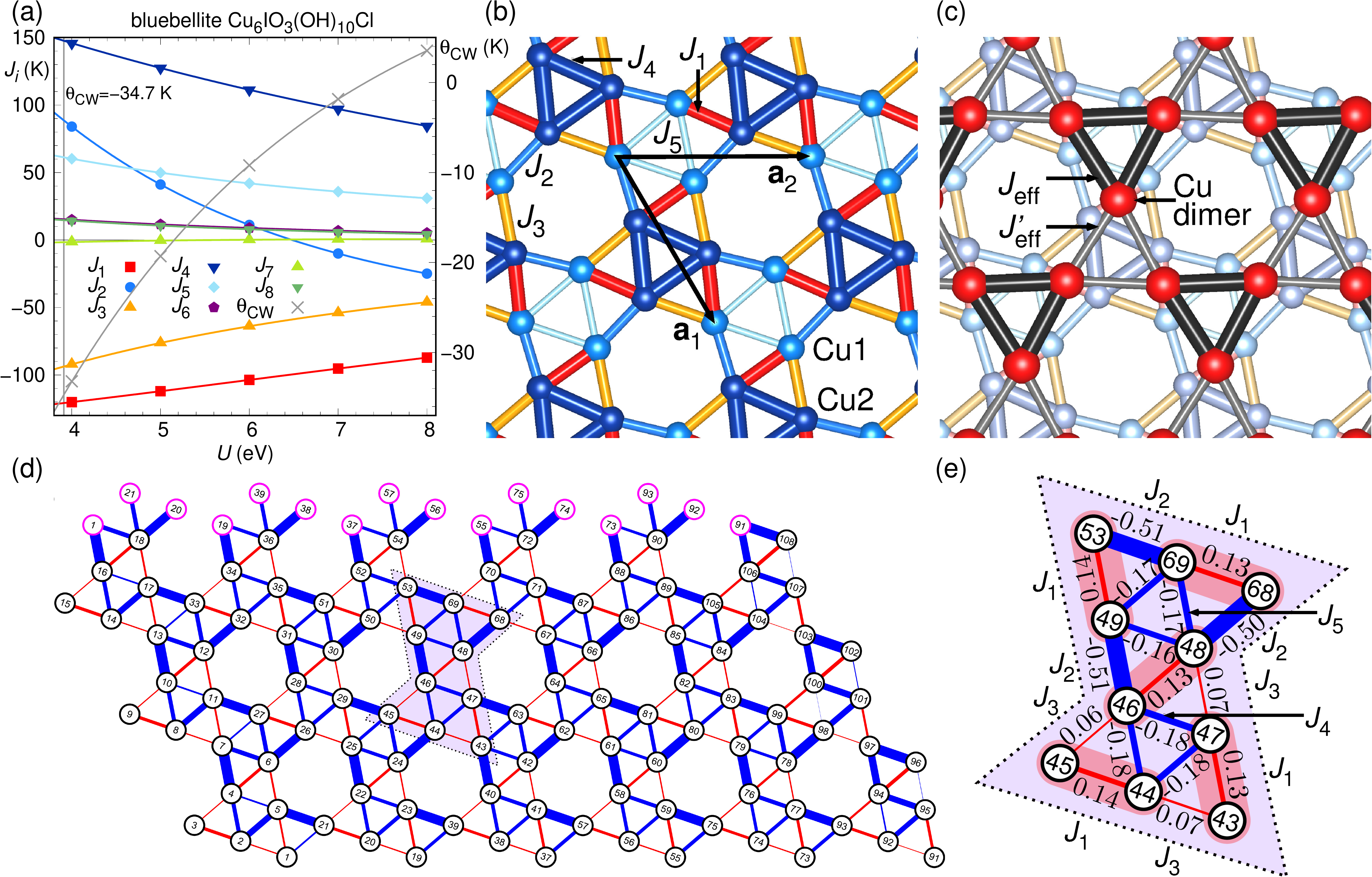}
    \caption{(a) DFT based energy mapping: First eight exchange interactions of bluebellite as function of on-site interaction strength $U$, at fixed $J_{\rm H}=1$\,eV. The vertical line indicates the $U$ value at which the Heisenberg Hamiltonian parameters yield the experimentally observed Curie-Weiss temperature~\cite{Haraguchi2021}. (b) bluebellite structure with the five "nearest neighbor" exchange paths defining the maple leaf lattice. The lattice vectors are given by $\mathbf{a}_1=\sqrt{7}/2(\hat{x}+\sqrt{3}\hat{y})$ and $\mathbf{a}_2=\sqrt{7}\hat{x}$. (c) Effective $S=1$ breathing kagome model with renormalized interactions. (d) Spin-spin correlations for all nearest-neighbor bonds obtained from DMRG on a 108 site maple-leaf cluster. The thickness of the bonds indicates the strength of the correlation and the color red (blue) indicates positive (negative) correlation. Note the clear dimerization in the ground state. (e) Enlargement from (d) with values of spin-spin correlations.
    }
    \label{fig1}
\end{figure*}

Here, we focus on bluebellite and resolve the most pressing issues for this prime example of a maple leaf antiferromagnet: we determine all relevant exchange interactions and solve the resulting Hamiltonian employing numerical and semi-analytical techniques. In particular, we address the question raised by experiment: why do the susceptibility and specific heat of bluebellite appear to have a Bonner-Fisher type shape, suggestive of one-dimensional systems? An answer to this question based on order-by-disorder was attempted without knowledge of the Hamiltonian~\cite{Makuta2021}. 
Methodologically, we apply the energy mapping technique which has proved valuable in extracting the Hamiltonian couplings for many important quantum spin systems~\cite{Jeschke2011,Jeschke2013,Guterding2016,Janson2016}. By virtue of a statistical approach and by extracting more than the apparently important exchange interactions, this method has led to surprising insights and met with much success for many materials~\cite{Iqbal2015,Iqbal2017,Iqbal2018,Ghosh2019,Chillal2020,Iida2020,Zivkovic2021}. We then perform density matrix renormalization group (DMRG) calculations~\cite{White1992} on the resulting maple leaf Hamiltonian. This method has been instrumental in furthering the comprehension of the physics of the kagome lattice antiferromagnet~\cite{He2017,Changlani-Spin-1_Kagome}. Our DMRG calculations show that bluebellite has a gapped valence bond solid ground state. In order to deepen our understanding, we develop a low-energy theory by implementing the standard bond operator formalism~\cite{BOT-Sachdev}. This theory permits one to perform calculations directly in the thermodynamic limit, obtain static and dynamical structure factors, and assess the behavior of thermodynamic quantities. The analytical and numerical calculations reveal that the bluebellite magnetic interactions very closely emulate an effective $S=1$ kagome Hamiltonian with a strong breathing anisotropy. So far, the possible $S=1$ kagome candidates, e.g., \ce{KV3Ge2O9}~\cite{Hara2012}, \ce{NaV6O11}~\cite{Kato2001}, $\mathrm{m\mbox{--}MPYNN \cdot BF_4}$~\cite{Wada1997}, all undergo lattice distortions at low temperatures. By establishing the connection between maple leaf and kagome, our work paves the pathway to possible realizations and synthesis of new effective $S=1$ kagome compounds emerging out of $S=1/2$ maple-leaf systems. As an experimental outlook, this enables the study of integer-spin highly frustrated kagome antiferromagnets, notably magnetization plateaus, excitations, and topological properties.   

\textit{Model Hamiltonian.-} Before we could extract the parameters of the Heisenberg Hamiltonian $H = \sum_{i<j} J_{ij} \mathbf{\hat S}_i \cdot \mathbf{\hat S}_j$ for bluebellite, we had to perform a relaxation of the internal positions of H, O and Cl while keeping Cu and I positions, as well as the lattice parameters fixed; this is necessary to sort out some obvious distortions in the experimentally determined room temperature structure~\cite{Haraguchi2021,SM}. Note, that the maple leaf lattice in the structure of synthetic bluebellite is much more regular compared to the structure of the mineral~\cite{Mills2014}. In applying the energy mapping technique, we make no assumptions as to which are the important exchange paths. Rather, we determine 20 couplings up to a Cu-Cu distance of 5.9\,{\AA} which is about twice the nearest neighbor distance. In Fig.~\ref{fig1}\,(a), we show how the first eight Heisenberg Hamiltonian parameters evolve with on-site Coulomb interaction $U$ applied to the strongly correlated Cu$^{2+}$ $3d$ orbitals. The relevant $U$ value is determined by calculating the Curie-Weiss temperature as $\theta_{\rm CW}=-\frac{1}{3}S(S+1)\sum_i z_i J_i$ where $z_i$ are the coordination numbers of $J_i$ and demanding that it matches the experimentally observed value $\theta_{\rm CW}=-34.7$\,K~\cite{Haraguchi2021}. 
The result for the five maple leaf couplings is $J_1 = -120.8(1.3)$\,K, $J_2 = 88.6(1.2)$\,K, $J_3 = -93.7(1.0)$\,K, $J_4 = 147.6(1.3)$\,K, $J_5 = 61.3(7)$\,K. At first glance, it is disappointing that two of the five couplings are ferromagnetic, making this Hamiltonian very distant from the ideal maple leaf antiferromagnet~\cite{Ghosh2022}. But the mixed ferro- and antiferromagnetic (FM and AFM) bluebellite Hamiltonian may attain very attractive properties. First of all, the strongest AFM $J_4$ would try to enforce a $120^\circ$ order on the $J_4$ triangles. However, it might not be sufficient to establish a full magnetic order on the system, as the $J_5$ bonds, which forms the other set of triangles on the lattice, is the weakest. On the other hand, the second strongest AFM $J_2$ bonds form hexagons of alternating interactions with the FM $J_3$ bonds, which might promote a dimerized singlet state on the hexagons. Finally, the strong FM $J_1$ bond might try to project an effective spin-$1$ on them and reduce the quantum fluctuations of the system. The complex interplay of all these effects makes the properties of bluebellite extremely intricate and intriguing, which we attempt to understand in the rest of this communication.

\textit{Ground State.-} 
First, we study the ground state of bluebellite with DMRG by using the iTENSOR library~\cite{itensor}. On three distinct lattice sizes—$48$, $108$, and $192$ site clusters—we conduct $24$–$30$ sweeps while maintaining a maximum bond dimension of $2048$~\cite{SM}. Based on finite-size-scaling, we reveal a magnetically disordered ground state with an energy per site $E_0/\widetilde{J}= -0.217(1)$, which is gapped by $\Delta/\widetilde{J}=0.106(2)$ ($\widetilde{J}=\sqrt{J_1^2+J_2^2+J_3^2+J_4^2+J_5^2}=238\text{ K}$ set as the energy scale)~\cite{SM}. We show the spin-spin correlations, $\langle\vec{S}_i\cdot\vec{S}_j\rangle$, for all nearest-neighbor spin pairs, $\vec{S}_i$ and $\vec{S}_j$, in Fig.~\ref{fig1}\,(d), which shows strong singlet formation in the spin pairs across the AFM $J_2$ bonds. This dimerization propensity on the $J_2$ bonds is indicative of a valence bond solid (VBS) state. 

For such a dimerized system, further critical insights can be obtained from the bond operator formalism~\cite{BOT-Sachdev}, which provides a suitable way to construct an effective low-energy bosonic theory in the thermodynamic limit. In bond operator representation, one transforms the spin basis to a dimer basis by writing the singlet and three triplet states on a bond as bosons (deemed as bond operators). In our analysis, the $J_2$ bonds define the dimers. We assume a singlet background (a product state of singlets on the $J_2$ bonds) as a mean-field, and the triplon (dispersing triplet) excitations on the singlet are treated systematically while ignoring any triplon-triplon interaction (details in the Supplemental Material~\cite{SM} which contains additional Refs.~\cite{Lyons1960,Kaplan2007,Brown2006}). This method has been successful in describing the ground state as well as thermodynamic properties of several magnetic materials~\cite{Normand2011,Plumb_2015,Kwon2001,Ghosh_Hida_Model_of_Kagome,Adhikary2021}. The results from this bosonic theory agree well with the DMRG calculations. It finds a stable VBS ground state with energy $E_0/\widetilde{J}= -0.22494$ and a spin gap of $\Delta/\widetilde{J}=0.13474$. The spin-spin correlations on all the bonds also agree with the DMRG results; the strongest, on the $J_2$ bonds, for example, is found to be $-0.57570$ (compare with Fig. \ref{fig1}\,(e)).

Apart from the strong singlets forming on the AFM $J_2$ bonds, the spins connected by FM $J_1$ interactions develop strong ferromagnetic correlations. The total spin moment on these FM bonds is estimated to be approximately $1.8$ from both methods, signalling the system's tendency to satisfy the strongest FM bond $J_1$ by forming triplets so that the $S=1/2$ astride the $J_1$ bonds would project onto an effective $S=1$ (for a full $S=1$ projection, the total spin moment $S(S+1)=2$ on a bond). Thus, we discover a simple but elegant property of the bluebellite Heisenberg Hamiltonian -- it emerges as an effective $S=1$ kagome system (see Fig.~\ref{fig1}\,(c)). This effective kagome lattice Hamiltonian will have two distinct first neighbor interactions for smaller triangles $J_{\rm eff}\propto J_2+J_5$, and for slightly larger triangles $J'_{\rm eff}\propto J_3+J_4$, leading to a breathing anisotropy (see Fig.~\ref{fig1}\,(b) and (e))~\cite{Iqbal-2018_breathing}. We have precisely evaluated the interactions in the effective $S=1$ breathing kagome lattice by repeating the energy mapping with $S=1/2$ moments connected by $J_1$ bonds constrained to $S=1$ (for details, see Ref.~\cite{SM}). We find antiferromagnetic couplings $J_{\rm eff}=49(2)$\,K and $J'_{\rm eff}=18(2)$\,K, and the breathing anisotropy is $J_{\rm eff}/J'_{\rm eff}\approx 2.7$ (Fig.~\ref{fig1}\,(c)).  

An analysis of the $S=1$ breathing kagome AFM due to Ghosh \textit{et. al.} revealed that the ground state of such a system is a trimerized singlet~\cite{Ghosh-Spin-1_Kagome}. To understand this ground state one can start with isolated $J_{\rm eff}$ triangles (see Fig.~\ref{fig1}\,(c)), i.e., set $J'_{\rm eff}$ to zero. In this limit, the ground state is a product of pure singlets on each \emph{down} triangle. When the $J'_{\rm eff}$ is turned on, triplet fluctuations develop on this singlet, which for $J'_{\rm eff}\le J_{\rm eff}$ remain insufficient to conjure a phase transition, forming a stable trimerized singlet ground state for the system~\cite{Ghosh-Spin-1_Kagome,Changlani-Spin-1_Kagome,Liu2015}. In the present case, a similar `trimerized' state is also realized, which translates as a dimerized state. If one tunes $J_1\rightarrow\infty$, this dimerized state on the maple-leaf lattice should continuously evolve to the trimerized singlet state of $S=1$ breathing kagome. The trimerized state here is akin to the valence-bond solid state for the AFM spin-$1$ chain due to Affleck, Kennedy, Lieb and Tasaki~\cite{AKLT}. In both cases, one sees the spin-$1$ as a combination of two spin-$1/2$, tries to satisfy the AFM interactions locally, and then projects a spin-$1$ out of two spin-$1/2$. Also note that for an AFM spin-$1$ chain with coupling equal to $J_2$, the spin-gap is $0.15$~\cite{Haldane1983}, which is slightly larger compared to the spin gap we obtain. A similar physics was seen for $S=1$ kagome in Ref.~\cite{Ghosh_Hida_Model_of_Kagome}. 

\begin{figure}[htb]
    \centering
    \includegraphics[width=\columnwidth]{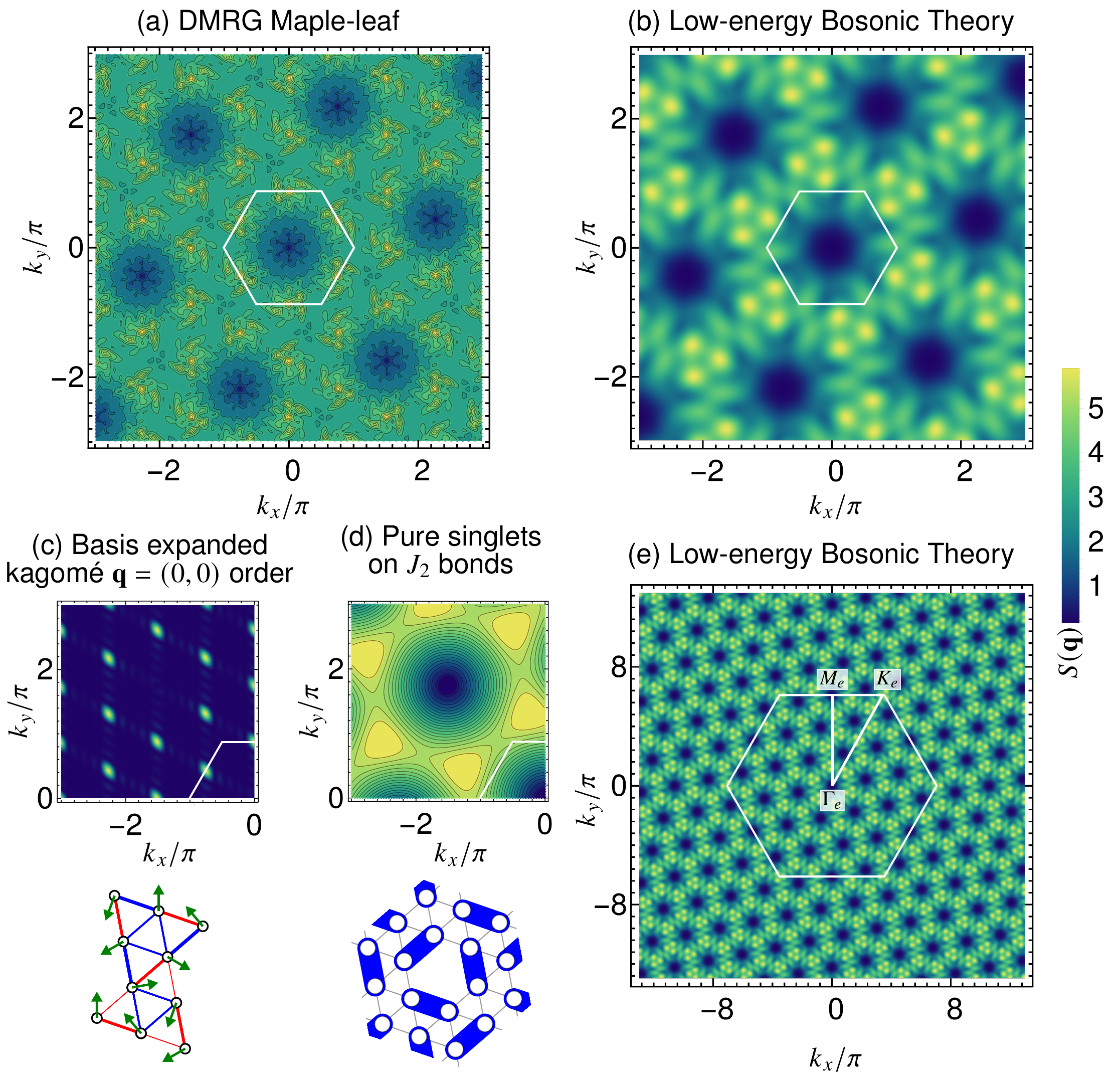}
    \caption{Static spin structure factor, $S_\mathbf{q}$, (a) obtained from DMRG on a 192 site maple-leaf cluster, (b) and (e) obtained from the bond-operator mean-field theory at the thermodynamic limit. (c) The $S_\mathbf{q}$ for the maple-leaf lattice obtained by doing a basis expansion of the $\mathbf{q}=(0,0)$ order on the classical kagome system. We also introduce a small canting between the spins on the $J_1$ bonds (see the bottom panel). We find Bragg peaks on the $M$ points of the fourth BZ of the system.  (d) The $S_\mathbf{q}$ for product state of pure singlets forming on the $J_2$ bonds. The maxima shifts significantly away from the $M$ points. Using the notation same as Fig. \ref{fig1}\,(d), the bottom panels of both (c) and (d) depicts the NN spin-spin correlations for corresponding states.  
    }
    \label{fig2}
\end{figure}

\textit{Static and Dynamical Structure Factors.-} 
We first calculate the static structure factor, $S(\mathbf{q})=\frac{1}{N}\sum_{ij}e^{\imath\mathbf{q}\cdot(\mathbf{r}_i-\mathbf{r}_j)}\langle\mathbf{\hat S}_i\cdot\mathbf{\hat S}_j\rangle$, using DMRG and bosonic theory. The DMRG result (Fig.~\ref{fig2}\,(a)), shows diffused peaks at the $M$ points of the fourth Brillouin zone (BZ). In contrast, the peaks from the low-energy bosonic theory (Fig.~\ref{fig2}\,(b)) are significantly shifted from the $M$ points. To understand this further, we exploit the relation between the kagome lattice and the maple-leaf lattice (see Fig.~\ref{fig1}\,(c)), to transform the static spin structure factor of kagome $\mathbf{q}=0$ order into the $S_\mathbf{q}$ for a maple-leaf system (see Ref.~\cite{SM}) and find Bragg peaks at the $M$ points (see Fig.~\ref{fig2}\,(c)). We also introduce a relative spin canting across $J_1$ bonds (see Fig.~\ref{fig2}\,(c)) and find it to be incapable of changing the Bragg peak positions. Therefore, a shift in the peaks in Fig.~\ref{fig2}\,(b) must have a quantum mechanical origin. To confirm this, we assume a state which is a product state of pure singlets on the $J_2$ bonds. Note that this state is similar to the mean-field wavefunction for our bosonic theory. This state reproduces the shift in question (see Fig.~\ref{fig2}\,(d)). Thus, we infer that the shift in the maximum of $S_\mathbf{q}$ away from the $M$ points is due to the stabilization of the spin-singlets on the $J_2$ bonds. The other modifications seen in Fig.~\ref{fig2}\,(b) are ascribed to triplet fluctuations. 

It is apparent from Fig.~\ref{fig2}\,(b), that a shift in the peak results in an enlargement of the BZ of the $S_\mathbf{q}$. A form factor calculation reveals that by traversing along either of the reciprocal vectors of the lattice, the structure factor is only periodic in seven reciprocal lattice spacing. Thus, the extended BZ of $S_\mathbf{q}$ is seven times larger than the actual BZ of the lattice (see Fig.~\ref{fig2}\,(e)). This significantly enhanced extended BZ is the reason behind the finite-size DMRG and bond-operator mean-field theory in the thermodynamic limit not agreeing in the calculation of $S(\mathbf{q})$.

\begin{figure}[b]
    \centering
    \includegraphics[width=\columnwidth]{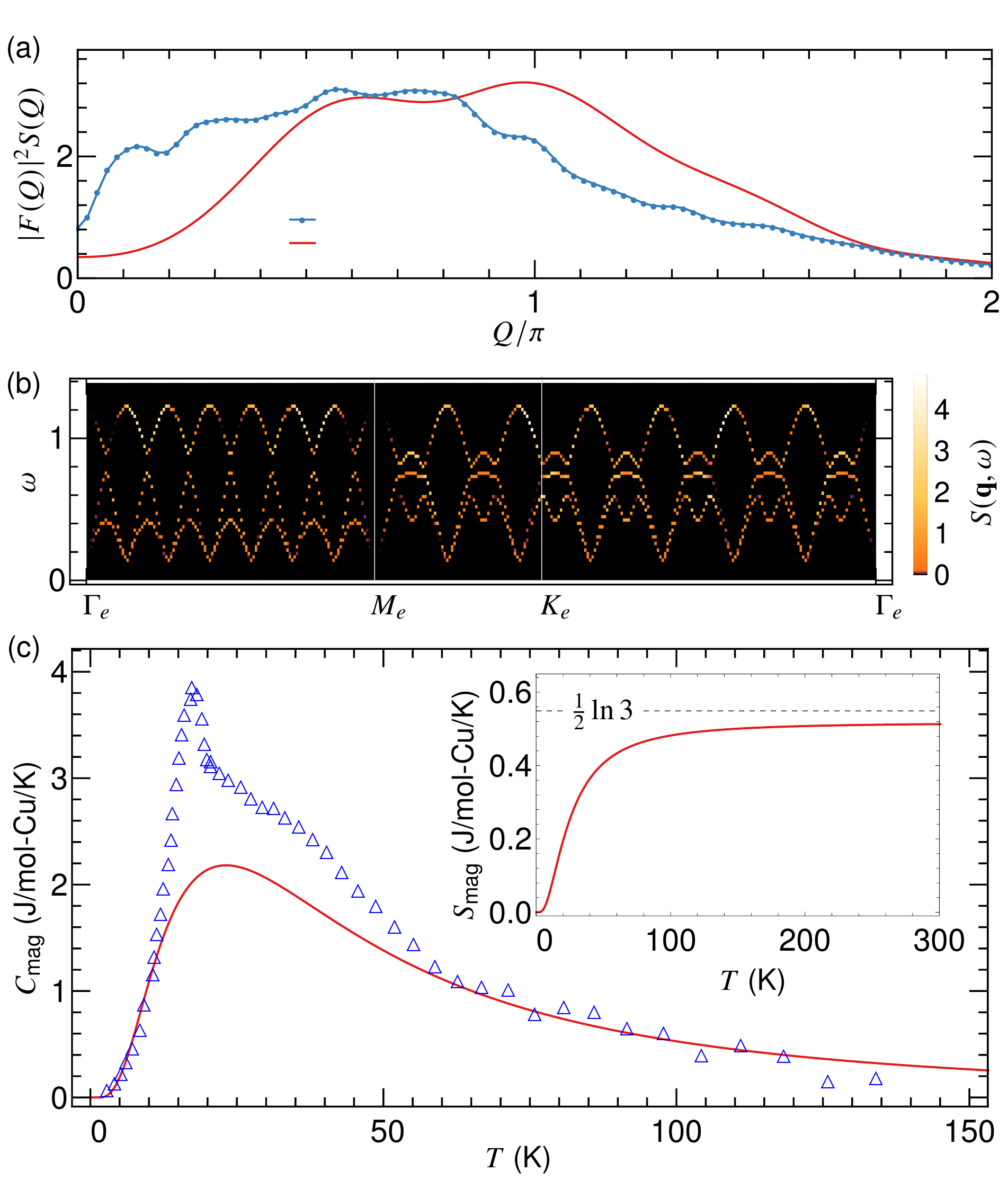}
    \caption{(a) Form factor modulated powder averaged static structure factor. (b) Dynamical spin structure factor calculated via bond-operator mean-field theory. (c) Magnetic specific heat $C_{\rm mag}$ obtained from bond-operator mean-field theory. Inset: Magnetic entropy $S_{\rm mag}$. }
    \label{fig3}
\end{figure}

We further calculate the powder averaged static spin structure factor, $
S(Q)=\frac{1}{4\pi} \int\mathrm{d}\Omega_{\hat{\mathbf{q}}}S(\mathbf{q})
$, with $Q=|\mathbf{q}|$, from both DMRG and the bond operator mean-field theory. We show the magnetic form factor, $F(Q)$, modulated $S(Q)$ in Fig.~\ref{fig3}\,(a). Apart from the static spin structure factor, we also calculate the dynamical spin structure factor, $
S(\mathbf{q},\omega)=\frac{1}{N}\sum_{ij}e^{\imath\mathbf{q}\cdot(\mathbf{r}_i-\mathbf{r}_j)}\int_{-\infty}^{\infty}\mathrm{d}t e^{\imath\omega t}\langle\mathbf{S}_j(t)\cdot\mathbf{S}_i(0)\rangle$,
for this system using the bosonic mean-field theory. We use the extended BZ of the system and plot $S(\mathbf{q},\omega)$ along the high-symmetry lines in Fig.~\ref{fig3}\,(b).  

\textit{Thermodynamic properties.-}
Capturing the finite temperature behavior precisely down to low temperature is a hard problem, but we can obtain qualitative information from our bond operator theory. We calculate specific heat using the bosonic mean-field theory (see \cite{SM}) and present it in Fig.~\ref{fig3}\,(c) together with the experimental result for the magnetic contribution to specific heat~\cite{Haraguchi2021}. It shows a broad peak around $22$\,K and matches the overall behavior of the experimental data. The specific heat shape that has been interpreted in terms of the Bonner-Fisher fit for the 1D Heisenberg chain~\cite{Haraguchi2021} is captured well by our entangled 2D wavefunction. This may be rationalized by the fact~\cite{Zheng2005} that zero-, one- and two-dimensional magnets can all have the same overall shape of specific heat, and magnetic susceptibility. The magnetic entropy calculation further validates effective spin-$1$ behavior of the system by approaching $\frac{1}{2}\ln{3}$ at high temperature (the factor of $1/2$ is due to two spin-$1/2$ combining to form a spin $1$), instead of $\ln{2}$ which would be the case for a fully AFM system. The residual entropy comes from the interlocking of the spins across the $J_1$ bonds at low temperature. 

However, the sharp peak seen in the experiment at $17$ K, which was identified as the onset of a magnetic order~\cite{Haraguchi2021}, is not seen in our computation, and we find a magnetically disordered ground state instead. The appearance of a magnetic order is indicative of substantial structural changes in bluebellite at low temperatures. However, we believe that this anomalous behavior can also be attributed to the thermal activation of other excited states which live above the triplet excitations~\cite{SM}. 

\textit{Conclusions.-}
We have determined the Heisenberg Hamiltonian for bluebellite by DFT energy mapping and found that two of the five couplings in the slightly distorted maple leaf lattice are ferromagnetic. Based on the DMRG result that the second largest antiferromagnetic exchange leads to very strong antiferromagnetic spin correlations, we have developed a bond operator mean field theory for bluebellite which gives us access to thermodynamic properties. Our calculated specific heat does not show a sharp ordering peak as the ground state for the Hamiltonian parameters determined for the room temperature crystal structure is magnetically disordered. However, there is excellent agreement in the overall shape of the specific heat, and we obtain the apparent Bonner-Fisher curve that was interpreted in terms of 1D physics~\cite{Haraguchi2021} from our fully two-dimensional maple leaf Hamiltonian. Furthermore, focusing on a pattern of strong ferromagnetic spin correlations in both DMRG and bond operator theory, we find that bluebellite realizes an effective spin-1 breathing kagome lattice. We predict both static and dynamic spin structure factors and show how they can be understood based on the classically expected $\mathbf{q}=\mathbf{0}$ order of the effective kagome model. We would like to point out that the maple leaf Hamiltonian and the effective spin-1 kagome model are determined from a room temperature crystal structure; they are both expected to show a 1/3 magnetization plateau which was not observed in the $T=4.2$\,K magnetization process~\cite{Haraguchi2021}. Therefore, it will be very interesting future work to determine a low temperature crystal structure of bluebellite experimentally and to study its Hamiltonian.

\acknowledgements

We thank Y. Haraguchi for allowing us to replot the magnetic specific heat. The work in W\"urzburg is supported by the Deutsche Forschungsgemeinschaft (DFG, German Research Foundation) through Project-ID 258499086-SFB 1170 and the Würzburg-Dresden Cluster of Excellence on Complexity and Topology in Quantum Matter – ct.qmat Project-ID 390858490-EXC 2147. R.\,T. and H.\,O.\,J. thank IIT Madras for a visiting faculty fellow position under the IoE program which facilitated the completion of this research work. Y.I. acknowledges financial support by the Science and Engineering Research Board (SERB), DST, India through the MATRICS Grant No.~MTR/2019/001042 and CEFIPRA Project No. 64T3-1. This research was supported in part by the National Science Foundation under Grant No.~NSF~PHY-1748958, ICTP through the Associates Programme and from the Simons Foundation through grant number 284558FY19, IIT Madras through the IoE program for establishing the QuCenDiEM group (Project No. SB20210813PHMHRD002720), the International Centre for Theoretical Sciences (ICTS), Bengaluru, India during a visit for participating in the program “Frustrated Metals and Insulators” (Code: ICTS/frumi2022/9), and the faculty associate program. Y.I. thanks IIT Madras for IoE travel grant to Germany which facilitated progress on the research work. Y.I. acknowledges the use of the computing resources at HPCE, IIT Madras.

\bibliography{bluebellite}

\begin{thebibliography}{56}%
\makeatletter
\providecommand \@ifxundefined [1]{%
 \@ifx{#1\undefined}
}%
\providecommand \@ifnum [1]{%
 \ifnum #1\expandafter \@firstoftwo
 \else \expandafter \@secondoftwo
 \fi
}%
\providecommand \@ifx [1]{%
 \ifx #1\expandafter \@firstoftwo
 \else \expandafter \@secondoftwo
 \fi
}%
\providecommand \natexlab [1]{#1}%
\providecommand \enquote  [1]{``#1''}%
\providecommand \bibnamefont  [1]{#1}%
\providecommand \bibfnamefont [1]{#1}%
\providecommand \citenamefont [1]{#1}%
\providecommand \href@noop [0]{\@secondoftwo}%
\providecommand \href [0]{\begingroup \@sanitize@url \@href}%
\providecommand \@href[1]{\@@startlink{#1}\@@href}%
\providecommand \@@href[1]{\endgroup#1\@@endlink}%
\providecommand \@sanitize@url [0]{\catcode `\\12\catcode `\$12\catcode
  `\&12\catcode `\#12\catcode `\^12\catcode `\_12\catcode `\%12\relax}%
\providecommand \@@startlink[1]{}%
\providecommand \@@endlink[0]{}%
\providecommand \url  [0]{\begingroup\@sanitize@url \@url }%
\providecommand \@url [1]{\endgroup\@href {#1}{\urlprefix }}%
\providecommand \urlprefix  [0]{URL }%
\providecommand \Eprint [0]{\href }%
\providecommand \doibase [0]{https://doi.org/}%
\providecommand \selectlanguage [0]{\@gobble}%
\providecommand \bibinfo  [0]{\@secondoftwo}%
\providecommand \bibfield  [0]{\@secondoftwo}%
\providecommand \translation [1]{[#1]}%
\providecommand \BibitemOpen [0]{}%
\providecommand \bibitemStop [0]{}%
\providecommand \bibitemNoStop [0]{.\EOS\space}%
\providecommand \EOS [0]{\spacefactor3000\relax}%
\providecommand \BibitemShut  [1]{\csname bibitem#1\endcsname}%
\let\auto@bib@innerbib\@empty
\bibitem [{\citenamefont {Balents}(2010)}]{Balents2010}%
  \BibitemOpen
  \bibfield  {author} {\bibinfo {author} {\bibfnamefont {L.}~\bibnamefont
  {Balents}},\ }\bibfield  {title} {\bibinfo {title} {Spin liquids in
  frustrated magnets},\ }\href {https://doi.org/10.1038/nature08917} {\bibfield
   {journal} {\bibinfo  {journal} {Nature}\ }\textbf {\bibinfo {volume}
  {464}},\ \bibinfo {pages} {199} (\bibinfo {year} {2010})}\BibitemShut
  {NoStop}%
\bibitem [{\citenamefont {Richter}\ \emph {et~al.}(2004)\citenamefont
  {Richter}, \citenamefont {Schulenburg},\ and\ \citenamefont
  {Honecker}}]{Richter2004}%
  \BibitemOpen
  \bibfield  {author} {\bibinfo {author} {\bibfnamefont {J.}~\bibnamefont
  {Richter}}, \bibinfo {author} {\bibfnamefont {J.}~\bibnamefont
  {Schulenburg}},\ and\ \bibinfo {author} {\bibfnamefont {A.}~\bibnamefont
  {Honecker}},\ }\bibinfo {title} {Quantum magnetism in two dimensions: From
  semi-classical n{\'e}el order to magnetic disorder},\ in\ \href
  {https://doi.org/10.1007/BFb0119592} {\emph {\bibinfo {booktitle} {Quantum
  Magnetism}}},\ \bibinfo {editor} {edited by\ \bibinfo {editor} {\bibfnamefont
  {U.}~\bibnamefont {Schollw{\"o}ck}}, \bibinfo {editor} {\bibfnamefont
  {J.}~\bibnamefont {Richter}}, \bibinfo {editor} {\bibfnamefont {D.~J.~J.}\
  \bibnamefont {Farnell}},\ and\ \bibinfo {editor} {\bibfnamefont {R.~F.}\
  \bibnamefont {Bishop}}}\ (\bibinfo  {publisher} {Springer Berlin
  Heidelberg},\ \bibinfo {address} {Berlin, Heidelberg},\ \bibinfo {year}
  {2004})\ pp.\ \bibinfo {pages} {85--153}\BibitemShut {NoStop}%
\bibitem [{\citenamefont {Lu}\ and\ \citenamefont {Kageyama}(2018)}]{Lu2018}%
  \BibitemOpen
  \bibfield  {author} {\bibinfo {author} {\bibfnamefont {H.}~\bibnamefont
  {Lu}}\ and\ \bibinfo {author} {\bibfnamefont {H.}~\bibnamefont {Kageyama}},\
  }\bibfield  {title} {\bibinfo {title} {\ce{PbFePO4F2} with a 1/6th bond
  depleted triangular lattice},\ }\href {https://doi.org/10.1039/c8dt03802c}
  {\bibfield  {journal} {\bibinfo  {journal} {Dalton Trans.}\ }\textbf
  {\bibinfo {volume} {47}},\ \bibinfo {pages} {15303} (\bibinfo {year}
  {2018})}\BibitemShut {NoStop}%
\bibitem [{\citenamefont {Mendels}\ and\ \citenamefont
  {Wills}(2011)}]{Mendels2011}%
  \BibitemOpen
  \bibfield  {author} {\bibinfo {author} {\bibfnamefont {P.}~\bibnamefont
  {Mendels}}\ and\ \bibinfo {author} {\bibfnamefont {A.~S.}\ \bibnamefont
  {Wills}},\ }\bibinfo {title} {Kagom{\'e} antiferromagnets: Materials vs. spin
  liquid behaviors},\ in\ \href {https://doi.org/10.1007/978-3-642-10589-0_9}
  {\emph {\bibinfo {booktitle} {Introduction to Frustrated Magnetism:
  Materials, Experiments, Theory}}},\ \bibinfo {editor} {edited by\ \bibinfo
  {editor} {\bibfnamefont {C.}~\bibnamefont {Lacroix}}, \bibinfo {editor}
  {\bibfnamefont {P.}~\bibnamefont {Mendels}},\ and\ \bibinfo {editor}
  {\bibfnamefont {F.}~\bibnamefont {Mila}}}\ (\bibinfo  {publisher} {Springer
  Berlin Heidelberg},\ \bibinfo {address} {Berlin, Heidelberg},\ \bibinfo
  {year} {2011})\ pp.\ \bibinfo {pages} {207--238}\BibitemShut {NoStop}%
\bibitem [{\citenamefont {Betts}(1995)}]{Betts1995}%
  \BibitemOpen
  \bibfield  {author} {\bibinfo {author} {\bibfnamefont {D.~D.}\ \bibnamefont
  {Betts}},\ }\bibfield  {title} {\bibinfo {title} {A new two-dimensional
  lattice of coordination number five},\ }\href@noop {} {\bibfield  {journal}
  {\bibinfo  {journal} {Proc. N. S. Inst. Sci.}\ }\textbf {\bibinfo {volume}
  {40}},\ \bibinfo {pages} {95} (\bibinfo {year} {1995})}\BibitemShut {NoStop}%
\bibitem [{\citenamefont {Schmalfu\ss{}}\ \emph {et~al.}(2002)\citenamefont
  {Schmalfu\ss{}}, \citenamefont {Tomczak}, \citenamefont {Schulenburg},\ and\
  \citenamefont {Richter}}]{Schmalfuss2002}%
  \BibitemOpen
  \bibfield  {author} {\bibinfo {author} {\bibfnamefont {D.}~\bibnamefont
  {Schmalfu\ss{}}}, \bibinfo {author} {\bibfnamefont {P.}~\bibnamefont
  {Tomczak}}, \bibinfo {author} {\bibfnamefont {J.}~\bibnamefont
  {Schulenburg}},\ and\ \bibinfo {author} {\bibfnamefont {J.}~\bibnamefont
  {Richter}},\ }\bibfield  {title} {\bibinfo {title} {The spin-$\frac{1}{2}$
  {H}eisenberg antiferromagnet on a $\frac{1}{7}$-depleted triangular lattice:
  Ground-state properties},\ }\href
  {https://doi.org/10.1103/PhysRevB.65.224405} {\bibfield  {journal} {\bibinfo
  {journal} {Phys. Rev. B}\ }\textbf {\bibinfo {volume} {65}},\ \bibinfo
  {pages} {224405} (\bibinfo {year} {2002})}\BibitemShut {NoStop}%
\bibitem [{\citenamefont {Farnell}\ \emph {et~al.}(2018)\citenamefont
  {Farnell}, \citenamefont {G\"otze}, \citenamefont {Schulenburg},
  \citenamefont {Zinke}, \citenamefont {Bishop},\ and\ \citenamefont
  {Li}}]{Farnell2018}%
  \BibitemOpen
  \bibfield  {author} {\bibinfo {author} {\bibfnamefont {D.~J.~J.}\
  \bibnamefont {Farnell}}, \bibinfo {author} {\bibfnamefont {O.}~\bibnamefont
  {G\"otze}}, \bibinfo {author} {\bibfnamefont {J.}~\bibnamefont
  {Schulenburg}}, \bibinfo {author} {\bibfnamefont {R.}~\bibnamefont {Zinke}},
  \bibinfo {author} {\bibfnamefont {R.~F.}\ \bibnamefont {Bishop}},\ and\
  \bibinfo {author} {\bibfnamefont {P.~H.~Y.}\ \bibnamefont {Li}},\ }\bibfield
  {title} {\bibinfo {title} {Interplay between lattice topology, frustration,
  and spin quantum number in quantum antiferromagnets on {A}rchimedean
  lattices},\ }\href {https://doi.org/10.1103/PhysRevB.98.224402} {\bibfield
  {journal} {\bibinfo  {journal} {Phys. Rev. B}\ }\textbf {\bibinfo {volume}
  {98}},\ \bibinfo {pages} {224402} (\bibinfo {year} {2018})}\BibitemShut
  {NoStop}%
\bibitem [{\citenamefont {Ghosh}\ \emph {et~al.}(2022)\citenamefont {Ghosh},
  \citenamefont {M\"uller},\ and\ \citenamefont {Thomale}}]{Ghosh2022}%
  \BibitemOpen
  \bibfield  {author} {\bibinfo {author} {\bibfnamefont {P.}~\bibnamefont
  {Ghosh}}, \bibinfo {author} {\bibfnamefont {T.}~\bibnamefont {M\"uller}},\
  and\ \bibinfo {author} {\bibfnamefont {R.}~\bibnamefont {Thomale}},\
  }\bibfield  {title} {\bibinfo {title} {Another exact ground state of a
  two-dimensional quantum antiferromagnet},\ }\href
  {https://doi.org/10.1103/PhysRevB.105.L180412} {\bibfield  {journal}
  {\bibinfo  {journal} {Phys. Rev. B}\ }\textbf {\bibinfo {volume} {105}},\
  \bibinfo {pages} {L180412} (\bibinfo {year} {2022})}\BibitemShut {NoStop}%
\bibitem [{\citenamefont {Shastry}\ and\ \citenamefont
  {Sutherland}(1981)}]{Shastry1981}%
  \BibitemOpen
  \bibfield  {author} {\bibinfo {author} {\bibfnamefont {B.~S.}\ \bibnamefont
  {Shastry}}\ and\ \bibinfo {author} {\bibfnamefont {B.}~\bibnamefont
  {Sutherland}},\ }\bibfield  {title} {\bibinfo {title} {Exact ground state of
  a quantum mechanical antiferromagnet},\ }\href
  {https://doi.org/10.1016/0378-4363(81)90838-X} {\bibfield  {journal}
  {\bibinfo  {journal} {Physica B+C}\ }\textbf {\bibinfo {volume} {108}},\
  \bibinfo {pages} {1069} (\bibinfo {year} {1981})}\BibitemShut {NoStop}%
\bibitem [{\citenamefont {Jim{\'e}nez}\ \emph {et~al.}(2021)\citenamefont
  {Jim{\'e}nez}, \citenamefont {Crone}, \citenamefont {Fogh}, \citenamefont
  {Zayed}, \citenamefont {Lortz}, \citenamefont {Pomjakushina}, \citenamefont
  {Conder}, \citenamefont {L{\"a}uchli}, \citenamefont {Weber}, \citenamefont
  {Wessel}, \citenamefont {Honecker}, \citenamefont {Normand}, \citenamefont
  {R{\"u}egg}, \citenamefont {Corboz}, \citenamefont {R{\o}nnow},\ and\
  \citenamefont {Mila}}]{Jimenez2021}%
  \BibitemOpen
  \bibfield  {author} {\bibinfo {author} {\bibfnamefont {J.~L.}\ \bibnamefont
  {Jim{\'e}nez}}, \bibinfo {author} {\bibfnamefont {S.~P.~G.}\ \bibnamefont
  {Crone}}, \bibinfo {author} {\bibfnamefont {E.}~\bibnamefont {Fogh}},
  \bibinfo {author} {\bibfnamefont {M.~E.}\ \bibnamefont {Zayed}}, \bibinfo
  {author} {\bibfnamefont {R.}~\bibnamefont {Lortz}}, \bibinfo {author}
  {\bibfnamefont {E.}~\bibnamefont {Pomjakushina}}, \bibinfo {author}
  {\bibfnamefont {K.}~\bibnamefont {Conder}}, \bibinfo {author} {\bibfnamefont
  {A.~M.}\ \bibnamefont {L{\"a}uchli}}, \bibinfo {author} {\bibfnamefont
  {L.}~\bibnamefont {Weber}}, \bibinfo {author} {\bibfnamefont
  {S.}~\bibnamefont {Wessel}}, \bibinfo {author} {\bibfnamefont
  {A.}~\bibnamefont {Honecker}}, \bibinfo {author} {\bibfnamefont
  {B.}~\bibnamefont {Normand}}, \bibinfo {author} {\bibfnamefont
  {C.}~\bibnamefont {R{\"u}egg}}, \bibinfo {author} {\bibfnamefont
  {P.}~\bibnamefont {Corboz}}, \bibinfo {author} {\bibfnamefont {H.~M.}\
  \bibnamefont {R{\o}nnow}},\ and\ \bibinfo {author} {\bibfnamefont
  {F.}~\bibnamefont {Mila}},\ }\bibfield  {title} {\bibinfo {title} {A quantum
  magnetic analogue to the critical point of water},\ }\href
  {https://doi.org/10.1038/s41586-021-03411-8} {\bibfield  {journal} {\bibinfo
  {journal} {Nature}\ }\textbf {\bibinfo {volume} {592}},\ \bibinfo {pages}
  {370} (\bibinfo {year} {2021})}\BibitemShut {NoStop}%
\bibitem [{\citenamefont {Shi}\ \emph {et~al.}(2022)\citenamefont {Shi},
  \citenamefont {Dissanayake}, \citenamefont {Corboz}, \citenamefont
  {Steinhardt}, \citenamefont {Graf}, \citenamefont {Silevitch}, \citenamefont
  {Dabkowska}, \citenamefont {Rosenbaum}, \citenamefont {Mila},\ and\
  \citenamefont {Haravifard}}]{Shi2022}%
  \BibitemOpen
  \bibfield  {author} {\bibinfo {author} {\bibfnamefont {Z.}~\bibnamefont
  {Shi}}, \bibinfo {author} {\bibfnamefont {S.}~\bibnamefont {Dissanayake}},
  \bibinfo {author} {\bibfnamefont {P.}~\bibnamefont {Corboz}}, \bibinfo
  {author} {\bibfnamefont {W.}~\bibnamefont {Steinhardt}}, \bibinfo {author}
  {\bibfnamefont {D.}~\bibnamefont {Graf}}, \bibinfo {author} {\bibfnamefont
  {D.~M.}\ \bibnamefont {Silevitch}}, \bibinfo {author} {\bibfnamefont {H.~A.}\
  \bibnamefont {Dabkowska}}, \bibinfo {author} {\bibfnamefont {T.~F.}\
  \bibnamefont {Rosenbaum}}, \bibinfo {author} {\bibfnamefont {F.}~\bibnamefont
  {Mila}},\ and\ \bibinfo {author} {\bibfnamefont {S.}~\bibnamefont
  {Haravifard}},\ }\bibfield  {title} {\bibinfo {title} {Discovery of quantum
  phases in the {S}hastry-{S}utherland compound \ce{SrCu2(BO3)2} under extreme
  conditions of field and pressure},\ }\href
  {https://doi.org/10.1038/s41467-022-30036-w} {\bibfield  {journal} {\bibinfo
  {journal} {Nat. Commun.}\ }\textbf {\bibinfo {volume} {13}},\ \bibinfo
  {pages} {2301} (\bibinfo {year} {2022})}\BibitemShut {NoStop}%
\bibitem [{\citenamefont {Lee}\ \emph {et~al.}(2019)\citenamefont {Lee},
  \citenamefont {You}, \citenamefont {Sachdev},\ and\ \citenamefont
  {Vishwanath}}]{deconf-SSM}%
  \BibitemOpen
  \bibfield  {author} {\bibinfo {author} {\bibfnamefont {J.~Y.}\ \bibnamefont
  {Lee}}, \bibinfo {author} {\bibfnamefont {Y.-Z.}\ \bibnamefont {You}},
  \bibinfo {author} {\bibfnamefont {S.}~\bibnamefont {Sachdev}},\ and\ \bibinfo
  {author} {\bibfnamefont {A.}~\bibnamefont {Vishwanath}},\ }\bibfield  {title}
  {\bibinfo {title} {Signatures of a deconfined phase transition on the
  {S}hastry-{S}utherland lattice: Applications to quantum critical
  \ce{SrCu2(BO3)2}},\ }\href {https://doi.org/10.1103/PhysRevX.9.041037}
  {\bibfield  {journal} {\bibinfo  {journal} {Phys. Rev. X}\ }\textbf {\bibinfo
  {volume} {9}},\ \bibinfo {pages} {041037} (\bibinfo {year}
  {2019})}\BibitemShut {NoStop}%
\bibitem [{\citenamefont {Corboz}\ and\ \citenamefont
  {Mila}(2014)}]{Bound-state-SSM}%
  \BibitemOpen
  \bibfield  {author} {\bibinfo {author} {\bibfnamefont {P.}~\bibnamefont
  {Corboz}}\ and\ \bibinfo {author} {\bibfnamefont {F.}~\bibnamefont {Mila}},\
  }\bibfield  {title} {\bibinfo {title} {Crystals of bound states in the
  magnetization plateaus of the {S}hastry-{S}utherland model},\ }\href
  {https://doi.org/10.1103/PhysRevLett.112.147203} {\bibfield  {journal}
  {\bibinfo  {journal} {Phys. Rev. Lett.}\ }\textbf {\bibinfo {volume} {112}},\
  \bibinfo {pages} {147203} (\bibinfo {year} {2014})}\BibitemShut {NoStop}%
\bibitem [{\citenamefont {Norman}(2018)}]{Norman2018}%
  \BibitemOpen
  \bibfield  {author} {\bibinfo {author} {\bibfnamefont {M.}~\bibnamefont
  {Norman}},\ }\bibfield  {title} {\bibinfo {title} {Copper tellurium oxides
  – a playground for magnetism},\ }\href
  {https://doi.org/10.1016/j.jmmm.2017.11.006} {\bibfield  {journal} {\bibinfo
  {journal} {J. Mag. Mag. Mater.}\ }\textbf {\bibinfo {volume} {452}},\
  \bibinfo {pages} {507} (\bibinfo {year} {2018})}\BibitemShut {NoStop}%
\bibitem [{\citenamefont {Inosov}(2018)}]{Inosov2018}%
  \BibitemOpen
  \bibfield  {author} {\bibinfo {author} {\bibfnamefont {D.}~\bibnamefont
  {Inosov}},\ }\bibfield  {title} {\bibinfo {title} {Quantum magnetism in
  minerals},\ }\href {https://doi.org/10.1080/00018732.2018.1571986} {\bibfield
   {journal} {\bibinfo  {journal} {Adv. Phys.}\ }\textbf {\bibinfo {volume}
  {67}},\ \bibinfo {pages} {149} (\bibinfo {year} {2018})}\BibitemShut
  {NoStop}%
\bibitem [{\citenamefont {Hawthorne}\ \emph {et~al.}(1993)\citenamefont
  {Hawthorne}, \citenamefont {Kimata},\ and\ \citenamefont
  {Eby}}]{Hawthorne1993}%
  \BibitemOpen
  \bibfield  {author} {\bibinfo {author} {\bibfnamefont {F.~C.}\ \bibnamefont
  {Hawthorne}}, \bibinfo {author} {\bibfnamefont {M.}~\bibnamefont {Kimata}},\
  and\ \bibinfo {author} {\bibfnamefont {R.~K.}\ \bibnamefont {Eby}},\
  }\bibfield  {title} {\bibinfo {title} {{The crystal structure of spangolite,
  a complex copper sulfate sheet mineral}},\ }\href@noop {} {\bibfield
  {journal} {\bibinfo  {journal} {Am. Mineralog.}\ }\textbf {\bibinfo {volume}
  {78}},\ \bibinfo {pages} {649} (\bibinfo {year} {1993})}\BibitemShut
  {NoStop}%
\bibitem [{\citenamefont {Olmi}\ \emph {et~al.}(1995)\citenamefont {Olmi},
  \citenamefont {Sabelli},\ and\ \citenamefont {Trosti-Ferroni}}]{Olmi1995}%
  \BibitemOpen
  \bibfield  {author} {\bibinfo {author} {\bibfnamefont {F.}~\bibnamefont
  {Olmi}}, \bibinfo {author} {\bibfnamefont {C.}~\bibnamefont {Sabelli}},\ and\
  \bibinfo {author} {\bibfnamefont {R.}~\bibnamefont {Trosti-Ferroni}},\
  }\bibfield  {title} {\bibinfo {title} {The crystal structure of sabelliite},\
  }\href {https://doi.org/10.1127/ejm/7/6/1331} {\bibfield  {journal} {\bibinfo
   {journal} {Eur. J. Mineral.}\ }\textbf {\bibinfo {volume} {7}},\ \bibinfo
  {pages} {1331} (\bibinfo {year} {1995})}\BibitemShut {NoStop}%
\bibitem [{\citenamefont {Mills}\ \emph {et~al.}(2014)\citenamefont {Mills},
  \citenamefont {Kampf}, \citenamefont {Christy}, \citenamefont {Housley},
  \citenamefont {Rossman}, \citenamefont {Reynolds},\ and\ \citenamefont
  {Marty}}]{Mills2014}%
  \BibitemOpen
  \bibfield  {author} {\bibinfo {author} {\bibfnamefont {S.~J.}\ \bibnamefont
  {Mills}}, \bibinfo {author} {\bibfnamefont {A.~R.}\ \bibnamefont {Kampf}},
  \bibinfo {author} {\bibfnamefont {A.~G.}\ \bibnamefont {Christy}}, \bibinfo
  {author} {\bibfnamefont {R.~M.}\ \bibnamefont {Housley}}, \bibinfo {author}
  {\bibfnamefont {G.~R.}\ \bibnamefont {Rossman}}, \bibinfo {author}
  {\bibfnamefont {R.~E.}\ \bibnamefont {Reynolds}},\ and\ \bibinfo {author}
  {\bibfnamefont {J.}~\bibnamefont {Marty}},\ }\bibfield  {title} {\bibinfo
  {title} {Bluebellite and mojaveite, two new minerals from the central
  {M}ojave {D}esert, {C}alifornia, {USA}},\ }\href
  {https://doi.org/10.1180/minmag.2014.078.5.15} {\bibfield  {journal}
  {\bibinfo  {journal} {Mineralog. Mag.}\ }\textbf {\bibinfo {volume} {78}},\
  \bibinfo {pages} {1325–1340} (\bibinfo {year} {2014})}\BibitemShut
  {NoStop}%
\bibitem [{\citenamefont {Kampf}\ \emph {et~al.}(2013)\citenamefont {Kampf},
  \citenamefont {Mills}, \citenamefont {Housley},\ and\ \citenamefont
  {Marty}}]{Kampf2013}%
  \BibitemOpen
  \bibfield  {author} {\bibinfo {author} {\bibfnamefont {A.~R.}\ \bibnamefont
  {Kampf}}, \bibinfo {author} {\bibfnamefont {S.~J.}\ \bibnamefont {Mills}},
  \bibinfo {author} {\bibfnamefont {R.~M.}\ \bibnamefont {Housley}},\ and\
  \bibinfo {author} {\bibfnamefont {J.}~\bibnamefont {Marty}},\ }\bibfield
  {title} {\bibinfo {title} {Lead-tellurium oxysalts from {O}tto {M}ountain
  near {B}aker, {C}alifornia: {VIII}. fuettererite, \ce{Pb3Cu^{2+}6
  Te^{6+}O6(OH)7Cl5}, a new mineral with double spangolite-type sheets},\
  }\href {https://doi.org/10.2138/am.2013.4218} {\bibfield  {journal} {\bibinfo
   {journal} {Am. Mineralog.}\ }\textbf {\bibinfo {volume} {98}},\ \bibinfo
  {pages} {506} (\bibinfo {year} {2013})}\BibitemShut {NoStop}%
\bibitem [{\citenamefont {Fennell}\ \emph {et~al.}(2011)\citenamefont
  {Fennell}, \citenamefont {Piatek}, \citenamefont {Stephenson}, \citenamefont
  {Nilsen},\ and\ \citenamefont {Rønnow}}]{Fennell2011}%
  \BibitemOpen
  \bibfield  {author} {\bibinfo {author} {\bibfnamefont {T.}~\bibnamefont
  {Fennell}}, \bibinfo {author} {\bibfnamefont {J.~O.}\ \bibnamefont {Piatek}},
  \bibinfo {author} {\bibfnamefont {R.~A.}\ \bibnamefont {Stephenson}},
  \bibinfo {author} {\bibfnamefont {G.~J.}\ \bibnamefont {Nilsen}},\ and\
  \bibinfo {author} {\bibfnamefont {H.~M.}\ \bibnamefont {Rønnow}},\
  }\bibfield  {title} {\bibinfo {title} {Spangolite: an s = 1/2 maple leaf
  lattice antiferromagnet?},\ }\href
  {https://doi.org/10.1088/0953-8984/23/16/164201} {\bibfield  {journal}
  {\bibinfo  {journal} {J. Phys.: Condens. Matter}\ }\textbf {\bibinfo {volume}
  {23}},\ \bibinfo {pages} {164201} (\bibinfo {year} {2011})}\BibitemShut
  {NoStop}%
\bibitem [{\citenamefont {Haraguchi}\ \emph {et~al.}(2021)\citenamefont
  {Haraguchi}, \citenamefont {Matsuo}, \citenamefont {Kindo},\ and\
  \citenamefont {Hiroi}}]{Haraguchi2021}%
  \BibitemOpen
  \bibfield  {author} {\bibinfo {author} {\bibfnamefont {Y.}~\bibnamefont
  {Haraguchi}}, \bibinfo {author} {\bibfnamefont {A.}~\bibnamefont {Matsuo}},
  \bibinfo {author} {\bibfnamefont {K.}~\bibnamefont {Kindo}},\ and\ \bibinfo
  {author} {\bibfnamefont {Z.}~\bibnamefont {Hiroi}},\ }\bibfield  {title}
  {\bibinfo {title} {Quantum antiferromagnet bluebellite comprising a
  maple-leaf lattice made of spin-1/2 {C}u$^{2+}$ ions},\ }\href
  {https://doi.org/10.1103/PhysRevB.104.174439} {\bibfield  {journal} {\bibinfo
   {journal} {Phys. Rev. B}\ }\textbf {\bibinfo {volume} {104}},\ \bibinfo
  {pages} {174439} (\bibinfo {year} {2021})}\BibitemShut {NoStop}%
\bibitem [{\citenamefont {Makuta}\ and\ \citenamefont
  {Hotta}(2021)}]{Makuta2021}%
  \BibitemOpen
  \bibfield  {author} {\bibinfo {author} {\bibfnamefont {R.}~\bibnamefont
  {Makuta}}\ and\ \bibinfo {author} {\bibfnamefont {C.}~\bibnamefont {Hotta}},\
  }\bibfield  {title} {\bibinfo {title} {Dimensional reduction in quantum
  spin-$\frac{1}{2}$ system on a $\frac{1}{7}$-depleted triangular lattice},\
  }\href {https://doi.org/10.1103/PhysRevB.104.224415} {\bibfield  {journal}
  {\bibinfo  {journal} {Phys. Rev. B}\ }\textbf {\bibinfo {volume} {104}},\
  \bibinfo {pages} {224415} (\bibinfo {year} {2021})}\BibitemShut {NoStop}%
\bibitem [{\citenamefont {Jeschke}\ \emph {et~al.}(2011)\citenamefont
  {Jeschke}, \citenamefont {Opahle}, \citenamefont {Kandpal}, \citenamefont
  {Valent\'{\i}}, \citenamefont {Das}, \citenamefont {Saha-Dasgupta},
  \citenamefont {Janson}, \citenamefont {Rosner}, \citenamefont {Br\"uhl},
  \citenamefont {Wolf}, \citenamefont {Lang}, \citenamefont {Richter},
  \citenamefont {Hu}, \citenamefont {Wang}, \citenamefont {Peters},
  \citenamefont {Pruschke},\ and\ \citenamefont {Honecker}}]{Jeschke2011}%
  \BibitemOpen
  \bibfield  {author} {\bibinfo {author} {\bibfnamefont {H.}~\bibnamefont
  {Jeschke}}, \bibinfo {author} {\bibfnamefont {I.}~\bibnamefont {Opahle}},
  \bibinfo {author} {\bibfnamefont {H.}~\bibnamefont {Kandpal}}, \bibinfo
  {author} {\bibfnamefont {R.}~\bibnamefont {Valent\'{\i}}}, \bibinfo {author}
  {\bibfnamefont {H.}~\bibnamefont {Das}}, \bibinfo {author} {\bibfnamefont
  {T.}~\bibnamefont {Saha-Dasgupta}}, \bibinfo {author} {\bibfnamefont
  {O.}~\bibnamefont {Janson}}, \bibinfo {author} {\bibfnamefont
  {H.}~\bibnamefont {Rosner}}, \bibinfo {author} {\bibfnamefont
  {A.}~\bibnamefont {Br\"uhl}}, \bibinfo {author} {\bibfnamefont
  {B.}~\bibnamefont {Wolf}}, \bibinfo {author} {\bibfnamefont {M.}~\bibnamefont
  {Lang}}, \bibinfo {author} {\bibfnamefont {J.}~\bibnamefont {Richter}},
  \bibinfo {author} {\bibfnamefont {S.}~\bibnamefont {Hu}}, \bibinfo {author}
  {\bibfnamefont {X.}~\bibnamefont {Wang}}, \bibinfo {author} {\bibfnamefont
  {R.}~\bibnamefont {Peters}}, \bibinfo {author} {\bibfnamefont
  {T.}~\bibnamefont {Pruschke}},\ and\ \bibinfo {author} {\bibfnamefont
  {A.}~\bibnamefont {Honecker}},\ }\bibfield  {title} {\bibinfo {title}
  {Multistep approach to microscopic models for frustrated quantum magnets: The
  case of the natural mineral azurite},\ }\href
  {https://doi.org/10.1103/PhysRevLett.106.217201} {\bibfield  {journal}
  {\bibinfo  {journal} {Phys. Rev. Lett.}\ }\textbf {\bibinfo {volume} {106}},\
  \bibinfo {pages} {217201} (\bibinfo {year} {2011})}\BibitemShut {NoStop}%
\bibitem [{\citenamefont {Jeschke}\ \emph {et~al.}(2013)\citenamefont
  {Jeschke}, \citenamefont {Salvat-Pujol},\ and\ \citenamefont
  {Valent\'{\i}}}]{Jeschke2013}%
  \BibitemOpen
  \bibfield  {author} {\bibinfo {author} {\bibfnamefont {H.~O.}\ \bibnamefont
  {Jeschke}}, \bibinfo {author} {\bibfnamefont {F.}~\bibnamefont
  {Salvat-Pujol}},\ and\ \bibinfo {author} {\bibfnamefont {R.}~\bibnamefont
  {Valent\'{\i}}},\ }\bibfield  {title} {\bibinfo {title} {First-principles
  determination of {H}eisenberg {H}amiltonian parameters for the
  spin-$\frac{1}{2}$ kagome antiferromagnet \ce{ZnCu3(OH)6Cl2}},\ }\href
  {https://doi.org/10.1103/PhysRevB.88.075106} {\bibfield  {journal} {\bibinfo
  {journal} {Phys. Rev. B}\ }\textbf {\bibinfo {volume} {88}},\ \bibinfo
  {pages} {075106} (\bibinfo {year} {2013})}\BibitemShut {NoStop}%
\bibitem [{\citenamefont {Guterding}\ \emph {et~al.}(2016)\citenamefont
  {Guterding}, \citenamefont {Valent\'{\i}},\ and\ \citenamefont
  {Jeschke}}]{Guterding2016}%
  \BibitemOpen
  \bibfield  {author} {\bibinfo {author} {\bibfnamefont {D.}~\bibnamefont
  {Guterding}}, \bibinfo {author} {\bibfnamefont {R.}~\bibnamefont
  {Valent\'{\i}}},\ and\ \bibinfo {author} {\bibfnamefont {H.~O.}\ \bibnamefont
  {Jeschke}},\ }\bibfield  {title} {\bibinfo {title} {Reduction of magnetic
  interlayer coupling in barlowite through isoelectronic substitution},\ }\href
  {https://doi.org/10.1103/PhysRevB.94.125136} {\bibfield  {journal} {\bibinfo
  {journal} {Phys. Rev. B}\ }\textbf {\bibinfo {volume} {94}},\ \bibinfo
  {pages} {125136} (\bibinfo {year} {2016})}\BibitemShut {NoStop}%
\bibitem [{\citenamefont {Janson}\ \emph {et~al.}(2016)\citenamefont {Janson},
  \citenamefont {Furukawa}, \citenamefont {Momoi}, \citenamefont {Sindzingre},
  \citenamefont {Richter},\ and\ \citenamefont {Held}}]{Janson2016}%
  \BibitemOpen
  \bibfield  {author} {\bibinfo {author} {\bibfnamefont {O.}~\bibnamefont
  {Janson}}, \bibinfo {author} {\bibfnamefont {S.}~\bibnamefont {Furukawa}},
  \bibinfo {author} {\bibfnamefont {T.}~\bibnamefont {Momoi}}, \bibinfo
  {author} {\bibfnamefont {P.}~\bibnamefont {Sindzingre}}, \bibinfo {author}
  {\bibfnamefont {J.}~\bibnamefont {Richter}},\ and\ \bibinfo {author}
  {\bibfnamefont {K.}~\bibnamefont {Held}},\ }\bibfield  {title} {\bibinfo
  {title} {Magnetic behavior of volborthite \ce{Cu3V2O7(OH)2.2H2O} determined
  by coupled trimers rather than frustrated chains},\ }\href
  {https://doi.org/10.1103/PhysRevLett.117.037206} {\bibfield  {journal}
  {\bibinfo  {journal} {Phys. Rev. Lett.}\ }\textbf {\bibinfo {volume} {117}},\
  \bibinfo {pages} {037206} (\bibinfo {year} {2016})}\BibitemShut {NoStop}%
\bibitem [{\citenamefont {Iqbal}\ \emph {et~al.}(2015)\citenamefont {Iqbal},
  \citenamefont {Jeschke}, \citenamefont {Reuther}, \citenamefont
  {Valent\'{\i}}, \citenamefont {Mazin}, \citenamefont {Greiter},\ and\
  \citenamefont {Thomale}}]{Iqbal2015}%
  \BibitemOpen
  \bibfield  {author} {\bibinfo {author} {\bibfnamefont {Y.}~\bibnamefont
  {Iqbal}}, \bibinfo {author} {\bibfnamefont {H.~O.}\ \bibnamefont {Jeschke}},
  \bibinfo {author} {\bibfnamefont {J.}~\bibnamefont {Reuther}}, \bibinfo
  {author} {\bibfnamefont {R.}~\bibnamefont {Valent\'{\i}}}, \bibinfo {author}
  {\bibfnamefont {I.~I.}\ \bibnamefont {Mazin}}, \bibinfo {author}
  {\bibfnamefont {M.}~\bibnamefont {Greiter}},\ and\ \bibinfo {author}
  {\bibfnamefont {R.}~\bibnamefont {Thomale}},\ }\bibfield  {title} {\bibinfo
  {title} {{Paramagnetism in the kagome compounds
  $(\mathrm{Zn},\mathrm{Mg},\mathrm{Cd}){\mathrm{Cu}}_{3}{(\mathrm{OH})}_{6}{\mathrm{Cl}}_{2}$}},\
  }\href {https://doi.org/10.1103/PhysRevB.92.220404} {\bibfield  {journal}
  {\bibinfo  {journal} {Phys. Rev. B}\ }\textbf {\bibinfo {volume} {92}},\
  \bibinfo {pages} {220404} (\bibinfo {year} {2015})}\BibitemShut {NoStop}%
\bibitem [{\citenamefont {Iqbal}\ \emph {et~al.}(2017)\citenamefont {Iqbal},
  \citenamefont {M\"uller}, \citenamefont {Riedl}, \citenamefont {Reuther},
  \citenamefont {Rachel}, \citenamefont {Valent\'{\i}}, \citenamefont
  {Gingras}, \citenamefont {Thomale},\ and\ \citenamefont
  {Jeschke}}]{Iqbal2017}%
  \BibitemOpen
  \bibfield  {author} {\bibinfo {author} {\bibfnamefont {Y.}~\bibnamefont
  {Iqbal}}, \bibinfo {author} {\bibfnamefont {T.}~\bibnamefont {M\"uller}},
  \bibinfo {author} {\bibfnamefont {K.}~\bibnamefont {Riedl}}, \bibinfo
  {author} {\bibfnamefont {J.}~\bibnamefont {Reuther}}, \bibinfo {author}
  {\bibfnamefont {S.}~\bibnamefont {Rachel}}, \bibinfo {author} {\bibfnamefont
  {R.}~\bibnamefont {Valent\'{\i}}}, \bibinfo {author} {\bibfnamefont
  {M.~J.~P.}\ \bibnamefont {Gingras}}, \bibinfo {author} {\bibfnamefont
  {R.}~\bibnamefont {Thomale}},\ and\ \bibinfo {author} {\bibfnamefont {H.~O.}\
  \bibnamefont {Jeschke}},\ }\bibfield  {title} {\bibinfo {title} {{Signatures
  of a gearwheel quantum spin liquid in a spin-$\frac{1}{2}$ pyrochlore
  molybdate Heisenberg antiferromagnet}},\ }\href
  {https://doi.org/10.1103/PhysRevMaterials.1.071201} {\bibfield  {journal}
  {\bibinfo  {journal} {Phys. Rev. Mater.}\ }\textbf {\bibinfo {volume} {1}},\
  \bibinfo {pages} {071201} (\bibinfo {year} {2017})}\BibitemShut {NoStop}%
\bibitem [{\citenamefont {Iqbal}\ \emph
  {et~al.}(2018{\natexlab{a}})\citenamefont {Iqbal}, \citenamefont {M\"uller},
  \citenamefont {Jeschke}, \citenamefont {Thomale},\ and\ \citenamefont
  {Reuther}}]{Iqbal2018}%
  \BibitemOpen
  \bibfield  {author} {\bibinfo {author} {\bibfnamefont {Y.}~\bibnamefont
  {Iqbal}}, \bibinfo {author} {\bibfnamefont {T.}~\bibnamefont {M\"uller}},
  \bibinfo {author} {\bibfnamefont {H.~O.}\ \bibnamefont {Jeschke}}, \bibinfo
  {author} {\bibfnamefont {R.}~\bibnamefont {Thomale}},\ and\ \bibinfo {author}
  {\bibfnamefont {J.}~\bibnamefont {Reuther}},\ }\bibfield  {title} {\bibinfo
  {title} {Stability of the spiral spin liquid in \ce{MnSc2S4}},\ }\href
  {https://doi.org/10.1103/PhysRevB.98.064427} {\bibfield  {journal} {\bibinfo
  {journal} {Phys. Rev. B}\ }\textbf {\bibinfo {volume} {98}},\ \bibinfo
  {pages} {064427} (\bibinfo {year} {2018}{\natexlab{a}})}\BibitemShut
  {NoStop}%
\bibitem [{\citenamefont {Ghosh}\ \emph {et~al.}(2019)\citenamefont {Ghosh},
  \citenamefont {Iqbal}, \citenamefont {M\"{u}ller}, \citenamefont
  {Ponnaganti}, \citenamefont {Thomale}, \citenamefont {Narayanan},
  \citenamefont {Reuther}, \citenamefont {Gingras},\ and\ \citenamefont
  {Jeschke}}]{Ghosh2019}%
  \BibitemOpen
  \bibfield  {author} {\bibinfo {author} {\bibfnamefont {P.}~\bibnamefont
  {Ghosh}}, \bibinfo {author} {\bibfnamefont {Y.}~\bibnamefont {Iqbal}},
  \bibinfo {author} {\bibfnamefont {T.}~\bibnamefont {M\"{u}ller}}, \bibinfo
  {author} {\bibfnamefont {R.~T.}\ \bibnamefont {Ponnaganti}}, \bibinfo
  {author} {\bibfnamefont {R.}~\bibnamefont {Thomale}}, \bibinfo {author}
  {\bibfnamefont {R.}~\bibnamefont {Narayanan}}, \bibinfo {author}
  {\bibfnamefont {J.}~\bibnamefont {Reuther}}, \bibinfo {author} {\bibfnamefont
  {M.~J.~P.}\ \bibnamefont {Gingras}},\ and\ \bibinfo {author} {\bibfnamefont
  {H.~O.}\ \bibnamefont {Jeschke}},\ }\bibfield  {title} {\bibinfo {title}
  {Breathing chromium spinels: a showcase for a variety of pyrochlore
  {H}eisenberg {H}amiltonians},\ }\href
  {https://doi.org/10.1038/s41535-019-0202-z} {\bibfield  {journal} {\bibinfo
  {journal} {npj Quantum Mater.}\ }\textbf {\bibinfo {volume} {4}},\ \bibinfo
  {pages} {63} (\bibinfo {year} {2019})}\BibitemShut {NoStop}%
\bibitem [{\citenamefont {Chillal}\ \emph {et~al.}(2020)\citenamefont
  {Chillal}, \citenamefont {Iqbal}, \citenamefont {Jeschke}, \citenamefont
  {Rodriguez-Rivera}, \citenamefont {Bewley}, \citenamefont {Manuel},
  \citenamefont {Khalyavin}, \citenamefont {Steffens}, \citenamefont {Thomale},
  \citenamefont {Islam}, \citenamefont {Reuther},\ and\ \citenamefont
  {Lake}}]{Chillal2020}%
  \BibitemOpen
  \bibfield  {author} {\bibinfo {author} {\bibfnamefont {S.}~\bibnamefont
  {Chillal}}, \bibinfo {author} {\bibfnamefont {Y.}~\bibnamefont {Iqbal}},
  \bibinfo {author} {\bibfnamefont {H.~O.}\ \bibnamefont {Jeschke}}, \bibinfo
  {author} {\bibfnamefont {J.~A.}\ \bibnamefont {Rodriguez-Rivera}}, \bibinfo
  {author} {\bibfnamefont {R.}~\bibnamefont {Bewley}}, \bibinfo {author}
  {\bibfnamefont {P.}~\bibnamefont {Manuel}}, \bibinfo {author} {\bibfnamefont
  {D.}~\bibnamefont {Khalyavin}}, \bibinfo {author} {\bibfnamefont
  {P.}~\bibnamefont {Steffens}}, \bibinfo {author} {\bibfnamefont
  {R.}~\bibnamefont {Thomale}}, \bibinfo {author} {\bibfnamefont {A.~T. M.~N.}\
  \bibnamefont {Islam}}, \bibinfo {author} {\bibfnamefont {J.}~\bibnamefont
  {Reuther}},\ and\ \bibinfo {author} {\bibfnamefont {B.}~\bibnamefont
  {Lake}},\ }\bibfield  {title} {\bibinfo {title} {Evidence for a
  three-dimensional quantum spin liquid in \ce{PbCuTe2O6}},\ }\href
  {https://doi.org/10.1038/s41467-020-15594-1} {\bibfield  {journal} {\bibinfo
  {journal} {Nat. Commun.}\ }\textbf {\bibinfo {volume} {11}},\ \bibinfo
  {pages} {2348} (\bibinfo {year} {2020})}\BibitemShut {NoStop}%
\bibitem [{\citenamefont {Iida}\ \emph {et~al.}(2020)\citenamefont {Iida},
  \citenamefont {Yoshida}, \citenamefont {Nakao}, \citenamefont {Jeschke},
  \citenamefont {Iqbal}, \citenamefont {Nakajima}, \citenamefont
  {Ohira-Kawamura}, \citenamefont {Munakata}, \citenamefont {Inamura},
  \citenamefont {Murai}, \citenamefont {Ishikado}, \citenamefont {Kumai},
  \citenamefont {Okada}, \citenamefont {Oda}, \citenamefont {Kakurai},\ and\
  \citenamefont {Matsuda}}]{Iida2020}%
  \BibitemOpen
  \bibfield  {author} {\bibinfo {author} {\bibfnamefont {K.}~\bibnamefont
  {Iida}}, \bibinfo {author} {\bibfnamefont {H.~K.}\ \bibnamefont {Yoshida}},
  \bibinfo {author} {\bibfnamefont {A.}~\bibnamefont {Nakao}}, \bibinfo
  {author} {\bibfnamefont {H.~O.}\ \bibnamefont {Jeschke}}, \bibinfo {author}
  {\bibfnamefont {Y.}~\bibnamefont {Iqbal}}, \bibinfo {author} {\bibfnamefont
  {K.}~\bibnamefont {Nakajima}}, \bibinfo {author} {\bibfnamefont
  {S.}~\bibnamefont {Ohira-Kawamura}}, \bibinfo {author} {\bibfnamefont
  {K.}~\bibnamefont {Munakata}}, \bibinfo {author} {\bibfnamefont
  {Y.}~\bibnamefont {Inamura}}, \bibinfo {author} {\bibfnamefont
  {N.}~\bibnamefont {Murai}}, \bibinfo {author} {\bibfnamefont
  {M.}~\bibnamefont {Ishikado}}, \bibinfo {author} {\bibfnamefont
  {R.}~\bibnamefont {Kumai}}, \bibinfo {author} {\bibfnamefont
  {T.}~\bibnamefont {Okada}}, \bibinfo {author} {\bibfnamefont
  {M.}~\bibnamefont {Oda}}, \bibinfo {author} {\bibfnamefont {K.}~\bibnamefont
  {Kakurai}},\ and\ \bibinfo {author} {\bibfnamefont {M.}~\bibnamefont
  {Matsuda}},\ }\bibfield  {title} {\bibinfo {title} {{$q=0$ long-range
  magnetic order in centennialite \ce{CaCu3(OD)6Cl2 \cdot {0.6}D2O}: A
  spin-$\frac{1}{2}$ perfect kagome antiferromagnet with
  ${J}_{1}\ensuremath{-}{J}_{2}\ensuremath{-}{J}_{d}$}},\ }\href
  {https://doi.org/10.1103/PhysRevB.101.220408} {\bibfield  {journal} {\bibinfo
   {journal} {Phys. Rev. B}\ }\textbf {\bibinfo {volume} {101}},\ \bibinfo
  {pages} {220408} (\bibinfo {year} {2020})}\BibitemShut {NoStop}%
\bibitem [{\citenamefont {\ifmmode \check{Z}\else
  \v{Z}\fi{}ivkovi\ifmmode~\acute{c}\else \'{c}\fi{}}\ \emph
  {et~al.}(2021)\citenamefont {\ifmmode \check{Z}\else
  \v{Z}\fi{}ivkovi\ifmmode~\acute{c}\else \'{c}\fi{}}, \citenamefont {Favre},
  \citenamefont {Salazar~Mejia}, \citenamefont {Jeschke}, \citenamefont
  {Magrez}, \citenamefont {Dabholkar}, \citenamefont {Noculak}, \citenamefont
  {Freitas}, \citenamefont {Jeong}, \citenamefont {Hegde}, \citenamefont
  {Testa}, \citenamefont {Babkevich}, \citenamefont {Su}, \citenamefont
  {Manuel}, \citenamefont {Luetkens}, \citenamefont {Baines}, \citenamefont
  {Baker}, \citenamefont {Wosnitza}, \citenamefont {Zaharko}, \citenamefont
  {Iqbal}, \citenamefont {Reuther},\ and\ \citenamefont
  {R\o{}nnow}}]{Zivkovic2021}%
  \BibitemOpen
  \bibfield  {author} {\bibinfo {author} {\bibfnamefont {I.}~\bibnamefont
  {\ifmmode \check{Z}\else \v{Z}\fi{}ivkovi\ifmmode~\acute{c}\else
  \'{c}\fi{}}}, \bibinfo {author} {\bibfnamefont {V.}~\bibnamefont {Favre}},
  \bibinfo {author} {\bibfnamefont {C.}~\bibnamefont {Salazar~Mejia}}, \bibinfo
  {author} {\bibfnamefont {H.~O.}\ \bibnamefont {Jeschke}}, \bibinfo {author}
  {\bibfnamefont {A.}~\bibnamefont {Magrez}}, \bibinfo {author} {\bibfnamefont
  {B.}~\bibnamefont {Dabholkar}}, \bibinfo {author} {\bibfnamefont
  {V.}~\bibnamefont {Noculak}}, \bibinfo {author} {\bibfnamefont {R.~S.}\
  \bibnamefont {Freitas}}, \bibinfo {author} {\bibfnamefont {M.}~\bibnamefont
  {Jeong}}, \bibinfo {author} {\bibfnamefont {N.~G.}\ \bibnamefont {Hegde}},
  \bibinfo {author} {\bibfnamefont {L.}~\bibnamefont {Testa}}, \bibinfo
  {author} {\bibfnamefont {P.}~\bibnamefont {Babkevich}}, \bibinfo {author}
  {\bibfnamefont {Y.}~\bibnamefont {Su}}, \bibinfo {author} {\bibfnamefont
  {P.}~\bibnamefont {Manuel}}, \bibinfo {author} {\bibfnamefont
  {H.}~\bibnamefont {Luetkens}}, \bibinfo {author} {\bibfnamefont
  {C.}~\bibnamefont {Baines}}, \bibinfo {author} {\bibfnamefont {P.~J.}\
  \bibnamefont {Baker}}, \bibinfo {author} {\bibfnamefont {J.}~\bibnamefont
  {Wosnitza}}, \bibinfo {author} {\bibfnamefont {O.}~\bibnamefont {Zaharko}},
  \bibinfo {author} {\bibfnamefont {Y.}~\bibnamefont {Iqbal}}, \bibinfo
  {author} {\bibfnamefont {J.}~\bibnamefont {Reuther}},\ and\ \bibinfo {author}
  {\bibfnamefont {H.~M.}\ \bibnamefont {R\o{}nnow}},\ }\bibfield  {title}
  {\bibinfo {title} {Magnetic field induced quantum spin liquid in the two
  coupled trillium lattices of \ce{K2Ni2(SO4)3}},\ }\href
  {https://doi.org/10.1103/PhysRevLett.127.157204} {\bibfield  {journal}
  {\bibinfo  {journal} {Phys. Rev. Lett.}\ }\textbf {\bibinfo {volume} {127}},\
  \bibinfo {pages} {157204} (\bibinfo {year} {2021})}\BibitemShut {NoStop}%
\bibitem [{\citenamefont {White}(1992)}]{White1992}%
  \BibitemOpen
  \bibfield  {author} {\bibinfo {author} {\bibfnamefont {S.~R.}\ \bibnamefont
  {White}},\ }\bibfield  {title} {\bibinfo {title} {Density matrix formulation
  for quantum renormalization groups},\ }\href
  {https://doi.org/10.1103/PhysRevLett.69.2863} {\bibfield  {journal} {\bibinfo
   {journal} {Phys. Rev. Lett.}\ }\textbf {\bibinfo {volume} {69}},\ \bibinfo
  {pages} {2863} (\bibinfo {year} {1992})}\BibitemShut {NoStop}%
\bibitem [{\citenamefont {He}\ \emph {et~al.}(2017)\citenamefont {He},
  \citenamefont {Zaletel}, \citenamefont {Oshikawa},\ and\ \citenamefont
  {Pollmann}}]{He2017}%
  \BibitemOpen
  \bibfield  {author} {\bibinfo {author} {\bibfnamefont {Y.-C.}\ \bibnamefont
  {He}}, \bibinfo {author} {\bibfnamefont {M.~P.}\ \bibnamefont {Zaletel}},
  \bibinfo {author} {\bibfnamefont {M.}~\bibnamefont {Oshikawa}},\ and\
  \bibinfo {author} {\bibfnamefont {F.}~\bibnamefont {Pollmann}},\ }\bibfield
  {title} {\bibinfo {title} {Signatures of dirac cones in a {DMRG} study of the
  kagome {H}eisenberg model},\ }\href
  {https://doi.org/10.1103/PhysRevX.7.031020} {\bibfield  {journal} {\bibinfo
  {journal} {Phys. Rev. X}\ }\textbf {\bibinfo {volume} {7}},\ \bibinfo {pages}
  {031020} (\bibinfo {year} {2017})}\BibitemShut {NoStop}%
\bibitem [{\citenamefont {Changlani}\ and\ \citenamefont
  {L\"auchli}(2015)}]{Changlani-Spin-1_Kagome}%
  \BibitemOpen
  \bibfield  {author} {\bibinfo {author} {\bibfnamefont {H.~J.}\ \bibnamefont
  {Changlani}}\ and\ \bibinfo {author} {\bibfnamefont {A.~M.}\ \bibnamefont
  {L\"auchli}},\ }\bibfield  {title} {\bibinfo {title} {Trimerized ground state
  of the spin-1 {H}eisenberg antiferromagnet on the kagome lattice},\ }\href
  {https://doi.org/10.1103/PhysRevB.91.100407} {\bibfield  {journal} {\bibinfo
  {journal} {Phys. Rev. B}\ }\textbf {\bibinfo {volume} {91}},\ \bibinfo
  {pages} {100407} (\bibinfo {year} {2015})}\BibitemShut {NoStop}%
\bibitem [{\citenamefont {Sachdev}\ and\ \citenamefont
  {Bhatt}(1990)}]{BOT-Sachdev}%
  \BibitemOpen
  \bibfield  {author} {\bibinfo {author} {\bibfnamefont {S.}~\bibnamefont
  {Sachdev}}\ and\ \bibinfo {author} {\bibfnamefont {R.~N.}\ \bibnamefont
  {Bhatt}},\ }\bibfield  {title} {\bibinfo {title} {Bond-operator
  representation of quantum spins: Mean-field theory of frustrated quantum
  {H}eisenberg antiferromagnets},\ }\href
  {https://doi.org/10.1103/PhysRevB.41.9323} {\bibfield  {journal} {\bibinfo
  {journal} {Phys. Rev. B}\ }\textbf {\bibinfo {volume} {41}},\ \bibinfo
  {pages} {9323} (\bibinfo {year} {1990})}\BibitemShut {NoStop}%
\bibitem [{\citenamefont {Hara}\ \emph {et~al.}(2012)\citenamefont {Hara},
  \citenamefont {Sato},\ and\ \citenamefont {Narumi}}]{Hara2012}%
  \BibitemOpen
  \bibfield  {author} {\bibinfo {author} {\bibfnamefont {S.}~\bibnamefont
  {Hara}}, \bibinfo {author} {\bibfnamefont {H.}~\bibnamefont {Sato}},\ and\
  \bibinfo {author} {\bibfnamefont {Y.}~\bibnamefont {Narumi}},\ }\bibfield
  {title} {\bibinfo {title} {Exotic magnetism of novel $s=1$ kagome lattice
  antiferromagnet \ce{KV3Ge2O9}},\ }\href
  {https://doi.org/10.1143/jpsj.81.073707} {\bibfield  {journal} {\bibinfo
  {journal} {J. Phys. Soc. Jpn}\ }\textbf {\bibinfo {volume} {81}},\ \bibinfo
  {pages} {073707} (\bibinfo {year} {2012})}\BibitemShut {NoStop}%
\bibitem [{\citenamefont {Kato}\ \emph {et~al.}(2001)\citenamefont {Kato},
  \citenamefont {Kato}, \citenamefont {Yoshimura},\ and\ \citenamefont
  {Kosuge}}]{Kato2001}%
  \BibitemOpen
  \bibfield  {author} {\bibinfo {author} {\bibfnamefont {H.}~\bibnamefont
  {Kato}}, \bibinfo {author} {\bibfnamefont {M.}~\bibnamefont {Kato}}, \bibinfo
  {author} {\bibfnamefont {K.}~\bibnamefont {Yoshimura}},\ and\ \bibinfo
  {author} {\bibfnamefont {K.}~\bibnamefont {Kosuge}},\ }\bibfield  {title}
  {\bibinfo {title} {${}^{23}${N}a-{NMR} study in \ce{NaV6O11}},\ }\href
  {https://doi.org/10.1143/jpsj.70.1404} {\bibfield  {journal} {\bibinfo
  {journal} {J. Phys. Soc. Jpn.}\ }\textbf {\bibinfo {volume} {70}},\ \bibinfo
  {pages} {1404} (\bibinfo {year} {2001})}\BibitemShut {NoStop}%
\bibitem [{\citenamefont {Wada}\ \emph {et~al.}(1997)\citenamefont {Wada},
  \citenamefont {Kobayashi}, \citenamefont {Yano}, \citenamefont {Okuno},
  \citenamefont {Yamaguchi},\ and\ \citenamefont {Awaga}}]{Wada1997}%
  \BibitemOpen
  \bibfield  {author} {\bibinfo {author} {\bibfnamefont {N.}~\bibnamefont
  {Wada}}, \bibinfo {author} {\bibfnamefont {T.}~\bibnamefont {Kobayashi}},
  \bibinfo {author} {\bibfnamefont {H.}~\bibnamefont {Yano}}, \bibinfo {author}
  {\bibfnamefont {T.}~\bibnamefont {Okuno}}, \bibinfo {author} {\bibfnamefont
  {A.}~\bibnamefont {Yamaguchi}},\ and\ \bibinfo {author} {\bibfnamefont
  {K.}~\bibnamefont {Awaga}},\ }\bibfield  {title} {\bibinfo {title}
  {Observation of spin-gap state in two-dimensional spin-1 kagomé
  antiferromagnet \ce{m-MPYNN\cdot BF4}},\ }\href
  {https://doi.org/10.1143/jpsj.66.961} {\bibfield  {journal} {\bibinfo
  {journal} {J. Phys. Soc. Jpn}\ }\textbf {\bibinfo {volume} {66}},\ \bibinfo
  {pages} {961} (\bibinfo {year} {1997})}\BibitemShut {NoStop}%
\bibitem [{SM()}]{SM}%
  \BibitemOpen
  \href@noop {} {}\bibinfo {note} {See Supplemental Material at { [TO BE
  INSERTED BY THE EDITORS]} for additional information on the structure
  preparation and energy mapping, bond operator mean-field theory details,
  additional DMRG results, the maple leaf structure factor and Luttinger-Tisza
  calculations.}\BibitemShut {Stop}%
\bibitem [{\citenamefont {Fishman}\ \emph {et~al.}(2022)\citenamefont
  {Fishman}, \citenamefont {White},\ and\ \citenamefont
  {Stoudenmire}}]{itensor}%
  \BibitemOpen
  \bibfield  {author} {\bibinfo {author} {\bibfnamefont {M.}~\bibnamefont
  {Fishman}}, \bibinfo {author} {\bibfnamefont {S.~R.}\ \bibnamefont {White}},\
  and\ \bibinfo {author} {\bibfnamefont {E.~M.}\ \bibnamefont {Stoudenmire}},\
  }\bibfield  {title} {\bibinfo {title} {{The ITensor Software Library for
  Tensor Network Calculations}},\ }\href
  {https://doi.org/10.21468/SciPostPhysCodeb.4} {\bibfield  {journal} {\bibinfo
   {journal} {SciPost Phys. Codebases}\ ,\ \bibinfo {pages} {4}} (\bibinfo
  {year} {2022})}\BibitemShut {NoStop}%
\bibitem [{\citenamefont {Lyons}\ and\ \citenamefont
  {Kaplan}(1960)}]{Lyons1960}%
  \BibitemOpen
  \bibfield  {author} {\bibinfo {author} {\bibfnamefont {D.~H.}\ \bibnamefont
  {Lyons}}\ and\ \bibinfo {author} {\bibfnamefont {T.~A.}\ \bibnamefont
  {Kaplan}},\ }\bibfield  {title} {\bibinfo {title} {{Method for determining
  ground-state spin configurations}},\ }\href
  {https://doi.org/10.1103/PhysRev.120.1580} {\bibfield  {journal} {\bibinfo
  {journal} {Phys. Rev.}\ }\textbf {\bibinfo {volume} {120}},\ \bibinfo {pages}
  {1580} (\bibinfo {year} {1960})}\BibitemShut {NoStop}%
\bibitem [{\citenamefont {Kaplan}\ and\ \citenamefont
  {Menyuk}(2007)}]{Kaplan2007}%
  \BibitemOpen
  \bibfield  {author} {\bibinfo {author} {\bibfnamefont {T.~A.}\ \bibnamefont
  {Kaplan}}\ and\ \bibinfo {author} {\bibfnamefont {N.}~\bibnamefont
  {Menyuk}},\ }\bibfield  {title} {\bibinfo {title} {{Spin ordering in
  three-dimensional crystals with strong competing exchange interactions}},\
  }\href {https://doi.org/10.1080/14786430601080229} {\bibfield  {journal}
  {\bibinfo  {journal} {Phil. Mag.}\ }\textbf {\bibinfo {volume} {87}},\
  \bibinfo {pages} {3711} (\bibinfo {year} {2007})}\BibitemShut {NoStop}%
\bibitem [{\citenamefont {Brown}\ \emph {et~al.}(2006)\citenamefont {Brown},
  \citenamefont {Fox}, \citenamefont {Maslen}, \citenamefont
  {O{\textquotesingle}Keefe},\ and\ \citenamefont {Willis}}]{Brown2006}%
  \BibitemOpen
  \bibfield  {author} {\bibinfo {author} {\bibfnamefont {P.~J.}\ \bibnamefont
  {Brown}}, \bibinfo {author} {\bibfnamefont {A.~G.}\ \bibnamefont {Fox}},
  \bibinfo {author} {\bibfnamefont {E.~N.}\ \bibnamefont {Maslen}}, \bibinfo
  {author} {\bibfnamefont {M.~A.}\ \bibnamefont {O{\textquotesingle}Keefe}},\
  and\ \bibinfo {author} {\bibfnamefont {B.~T.~M.}\ \bibnamefont {Willis}},\
  }\bibfield  {title} {\bibinfo {title} {Intensity of diffracted intensities},\
  }in\ \href {https://doi.org/10.1107/97809553602060000600} {\emph {\bibinfo
  {booktitle} {International Tables for Crystallography}}}\ (\bibinfo
  {publisher} {International Union of Crystallography},\ \bibinfo {year}
  {2006})\ pp.\ \bibinfo {pages} {554--595}\BibitemShut {NoStop}%
\bibitem [{\citenamefont {Normand}\ and\ \citenamefont
  {R\"uegg}(2011)}]{Normand2011}%
  \BibitemOpen
  \bibfield  {author} {\bibinfo {author} {\bibfnamefont {B.}~\bibnamefont
  {Normand}}\ and\ \bibinfo {author} {\bibfnamefont {C.}~\bibnamefont
  {R\"uegg}},\ }\bibfield  {title} {\bibinfo {title} {Complete bond-operator
  theory of the two-chain spin ladder},\ }\href
  {https://doi.org/10.1103/PhysRevB.83.054415} {\bibfield  {journal} {\bibinfo
  {journal} {Phys. Rev. B}\ }\textbf {\bibinfo {volume} {83}},\ \bibinfo
  {pages} {054415} (\bibinfo {year} {2011})}\BibitemShut {NoStop}%
\bibitem [{\citenamefont {Plumb}\ \emph {et~al.}(2015)\citenamefont {Plumb},
  \citenamefont {Hwang}, \citenamefont {Qiu}, \citenamefont {Harriger},
  \citenamefont {Granroth}, \citenamefont {Kolesnikov}, \citenamefont {Shu},
  \citenamefont {Chou}, \citenamefont {Rüegg}, \citenamefont {Kim},\ and\
  \citenamefont {Kim}}]{Plumb_2015}%
  \BibitemOpen
  \bibfield  {author} {\bibinfo {author} {\bibfnamefont {K.~W.}\ \bibnamefont
  {Plumb}}, \bibinfo {author} {\bibfnamefont {K.}~\bibnamefont {Hwang}},
  \bibinfo {author} {\bibfnamefont {Y.}~\bibnamefont {Qiu}}, \bibinfo {author}
  {\bibfnamefont {L.~W.}\ \bibnamefont {Harriger}}, \bibinfo {author}
  {\bibfnamefont {G.~E.}\ \bibnamefont {Granroth}}, \bibinfo {author}
  {\bibfnamefont {A.~I.}\ \bibnamefont {Kolesnikov}}, \bibinfo {author}
  {\bibfnamefont {G.~J.}\ \bibnamefont {Shu}}, \bibinfo {author} {\bibfnamefont
  {F.~C.}\ \bibnamefont {Chou}}, \bibinfo {author} {\bibfnamefont
  {C.}~\bibnamefont {Rüegg}}, \bibinfo {author} {\bibfnamefont {Y.~B.}\
  \bibnamefont {Kim}},\ and\ \bibinfo {author} {\bibfnamefont {Y.-J.}\
  \bibnamefont {Kim}},\ }\bibfield  {title} {\bibinfo {title}
  {Quasiparticle-continuum level repulsion in a quantum magnet},\ }\href
  {https://doi.org/10.1038/nphys3566} {\bibfield  {journal} {\bibinfo
  {journal} {Nat. Phys.}\ }\textbf {\bibinfo {volume} {12}},\ \bibinfo {pages}
  {224} (\bibinfo {year} {2015})}\BibitemShut {NoStop}%
\bibitem [{\citenamefont {Park}\ and\ \citenamefont
  {Sachdev}(2001)}]{Kwon2001}%
  \BibitemOpen
  \bibfield  {author} {\bibinfo {author} {\bibfnamefont {K.}~\bibnamefont
  {Park}}\ and\ \bibinfo {author} {\bibfnamefont {S.}~\bibnamefont {Sachdev}},\
  }\bibfield  {title} {\bibinfo {title} {Bond-operator theory of doped
  antiferromagnets: From mott insulators with bond-centered charge order to
  superconductors with nodal fermions},\ }\href
  {https://doi.org/10.1103/PhysRevB.64.184510} {\bibfield  {journal} {\bibinfo
  {journal} {Phys. Rev. B}\ }\textbf {\bibinfo {volume} {64}},\ \bibinfo
  {pages} {184510} (\bibinfo {year} {2001})}\BibitemShut {NoStop}%
\bibitem [{\citenamefont {Ghosh}\ and\ \citenamefont
  {Kumar}(2018)}]{Ghosh_Hida_Model_of_Kagome}%
  \BibitemOpen
  \bibfield  {author} {\bibinfo {author} {\bibfnamefont {P.}~\bibnamefont
  {Ghosh}}\ and\ \bibinfo {author} {\bibfnamefont {B.}~\bibnamefont {Kumar}},\
  }\bibfield  {title} {\bibinfo {title} {Spontaneous dimerization and moment
  formation in the {H}ida model of the spin-1 kagome antiferromagnet},\ }\href
  {https://doi.org/10.1103/PhysRevB.97.014413} {\bibfield  {journal} {\bibinfo
  {journal} {Phys. Rev. B}\ }\textbf {\bibinfo {volume} {97}},\ \bibinfo
  {pages} {014413} (\bibinfo {year} {2018})}\BibitemShut {NoStop}%
\bibitem [{\citenamefont {Adhikary}\ \emph {et~al.}(2021)\citenamefont
  {Adhikary}, \citenamefont {{R}alko},\ and\ \citenamefont
  {Kumar}}]{Adhikary2021}%
  \BibitemOpen
  \bibfield  {author} {\bibinfo {author} {\bibfnamefont {M.}~\bibnamefont
  {Adhikary}}, \bibinfo {author} {\bibfnamefont {A.}~\bibnamefont {{R}alko}},\
  and\ \bibinfo {author} {\bibfnamefont {B.}~\bibnamefont {Kumar}},\ }\bibfield
   {title} {\bibinfo {title} {Quantum paramagnetism and magnetization plateaus
  in a kagome-honeycomb {H}eisenberg antiferromagnet},\ }\href
  {https://doi.org/10.1103/PhysRevB.104.094416} {\bibfield  {journal} {\bibinfo
   {journal} {Phys. Rev. B}\ }\textbf {\bibinfo {volume} {104}},\ \bibinfo
  {pages} {094416} (\bibinfo {year} {2021})}\BibitemShut {NoStop}%
\bibitem [{\citenamefont {Iqbal}\ \emph
  {et~al.}(2018{\natexlab{b}})\citenamefont {Iqbal}, \citenamefont {Poilblanc},
  \citenamefont {Thomale},\ and\ \citenamefont {Becca}}]{Iqbal-2018_breathing}%
  \BibitemOpen
  \bibfield  {author} {\bibinfo {author} {\bibfnamefont {Y.}~\bibnamefont
  {Iqbal}}, \bibinfo {author} {\bibfnamefont {D.}~\bibnamefont {Poilblanc}},
  \bibinfo {author} {\bibfnamefont {R.}~\bibnamefont {Thomale}},\ and\ \bibinfo
  {author} {\bibfnamefont {F.}~\bibnamefont {Becca}},\ }\bibfield  {title}
  {\bibinfo {title} {{Persistence of the gapless spin liquid in the breathing
  kagome Heisenberg antiferromagnet}},\ }\href
  {https://doi.org/10.1103/PhysRevB.97.115127} {\bibfield  {journal} {\bibinfo
  {journal} {Phys. Rev. B}\ }\textbf {\bibinfo {volume} {97}},\ \bibinfo
  {pages} {115127} (\bibinfo {year} {2018}{\natexlab{b}})}\BibitemShut
  {NoStop}%
\bibitem [{\citenamefont {Ghosh}\ \emph {et~al.}(2016)\citenamefont {Ghosh},
  \citenamefont {Verma},\ and\ \citenamefont {Kumar}}]{Ghosh-Spin-1_Kagome}%
  \BibitemOpen
  \bibfield  {author} {\bibinfo {author} {\bibfnamefont {P.}~\bibnamefont
  {Ghosh}}, \bibinfo {author} {\bibfnamefont {A.~K.}\ \bibnamefont {Verma}},\
  and\ \bibinfo {author} {\bibfnamefont {B.}~\bibnamefont {Kumar}},\ }\bibfield
   {title} {\bibinfo {title} {Plaquette-triplon analysis of magnetic disorder
  and order in a trimerized spin-1 kagome {H}eisenberg antiferromagnet},\
  }\href {https://doi.org/10.1103/PhysRevB.93.014427} {\bibfield  {journal}
  {\bibinfo  {journal} {Phys. Rev. B}\ }\textbf {\bibinfo {volume} {93}},\
  \bibinfo {pages} {014427} (\bibinfo {year} {2016})}\BibitemShut {NoStop}%
\bibitem [{\citenamefont {Liu}\ \emph {et~al.}(2015)\citenamefont {Liu},
  \citenamefont {Li}, \citenamefont {Weichselbaum}, \citenamefont {von Delft},\
  and\ \citenamefont {Su}}]{Liu2015}%
  \BibitemOpen
  \bibfield  {author} {\bibinfo {author} {\bibfnamefont {T.}~\bibnamefont
  {Liu}}, \bibinfo {author} {\bibfnamefont {W.}~\bibnamefont {Li}}, \bibinfo
  {author} {\bibfnamefont {A.}~\bibnamefont {Weichselbaum}}, \bibinfo {author}
  {\bibfnamefont {J.}~\bibnamefont {von Delft}},\ and\ \bibinfo {author}
  {\bibfnamefont {G.}~\bibnamefont {Su}},\ }\bibfield  {title} {\bibinfo
  {title} {Simplex valence-bond crystal in the spin-1 kagome {H}eisenberg
  antiferromagnet},\ }\href {https://doi.org/10.1103/PhysRevB.91.060403}
  {\bibfield  {journal} {\bibinfo  {journal} {Phys. Rev. B}\ }\textbf {\bibinfo
  {volume} {91}},\ \bibinfo {pages} {060403} (\bibinfo {year}
  {2015})}\BibitemShut {NoStop}%
\bibitem [{\citenamefont {Affleck}\ \emph {et~al.}(1987)\citenamefont
  {Affleck}, \citenamefont {Kennedy}, \citenamefont {Lieb},\ and\ \citenamefont
  {Tasaki}}]{AKLT}%
  \BibitemOpen
  \bibfield  {author} {\bibinfo {author} {\bibfnamefont {I.}~\bibnamefont
  {Affleck}}, \bibinfo {author} {\bibfnamefont {T.}~\bibnamefont {Kennedy}},
  \bibinfo {author} {\bibfnamefont {E.~H.}\ \bibnamefont {Lieb}},\ and\
  \bibinfo {author} {\bibfnamefont {H.}~\bibnamefont {Tasaki}},\ }\bibfield
  {title} {\bibinfo {title} {Rigorous results on valence-bond ground states in
  antiferromagnets},\ }\href {https://doi.org/10.1103/PhysRevLett.59.799}
  {\bibfield  {journal} {\bibinfo  {journal} {Phys. Rev. Lett.}\ }\textbf
  {\bibinfo {volume} {59}},\ \bibinfo {pages} {799} (\bibinfo {year}
  {1987})}\BibitemShut {NoStop}%
\bibitem [{\citenamefont {Haldane}(1983)}]{Haldane1983}%
  \BibitemOpen
  \bibfield  {author} {\bibinfo {author} {\bibfnamefont {F.}~\bibnamefont
  {Haldane}},\ }\bibfield  {title} {\bibinfo {title} {{Continuum dynamics of
  the 1-D Heisenberg antiferromagnet: Identification with the O(3) nonlinear
  sigma model}},\ }\href {https://doi.org/10.1016/0375-9601(83)90631-x}
  {\bibfield  {journal} {\bibinfo  {journal} {Phys. Lett. A}\ }\textbf
  {\bibinfo {volume} {93}},\ \bibinfo {pages} {464} (\bibinfo {year}
  {1983})}\BibitemShut {NoStop}%
\bibitem [{\citenamefont {Zheng}\ \emph {et~al.}(2005)\citenamefont {Zheng},
  \citenamefont {Singh}, \citenamefont {McKenzie},\ and\ \citenamefont
  {Coldea}}]{Zheng2005}%
  \BibitemOpen
  \bibfield  {author} {\bibinfo {author} {\bibfnamefont {W.}~\bibnamefont
  {Zheng}}, \bibinfo {author} {\bibfnamefont {R.~R.~P.}\ \bibnamefont {Singh}},
  \bibinfo {author} {\bibfnamefont {R.~H.}\ \bibnamefont {McKenzie}},\ and\
  \bibinfo {author} {\bibfnamefont {R.}~\bibnamefont {Coldea}},\ }\bibfield
  {title} {\bibinfo {title} {Temperature dependence of the magnetic
  susceptibility for triangular-lattice antiferromagnets with spatially
  anisotropic exchange constants},\ }\href
  {https://doi.org/10.1103/PhysRevB.71.134422} {\bibfield  {journal} {\bibinfo
  {journal} {Phys. Rev. B}\ }\textbf {\bibinfo {volume} {71}},\ \bibinfo
  {pages} {134422} (\bibinfo {year} {2005})}\BibitemShut {NoStop}%
\end{thebibliography}%


\begin{thebibliography}{8}%
\makeatletter
\providecommand \@ifxundefined [1]{%
 \@ifx{#1\undefined}
}%
\providecommand \@ifnum [1]{%
 \ifnum #1\expandafter \@firstoftwo
 \else \expandafter \@secondoftwo
 \fi
}%
\providecommand \@ifx [1]{%
 \ifx #1\expandafter \@firstoftwo
 \else \expandafter \@secondoftwo
 \fi
}%
\providecommand \natexlab [1]{#1}%
\providecommand \enquote  [1]{``#1''}%
\providecommand \bibnamefont  [1]{#1}%
\providecommand \bibfnamefont [1]{#1}%
\providecommand \citenamefont [1]{#1}%
\providecommand \href@noop [0]{\@secondoftwo}%
\providecommand \href [0]{\begingroup \@sanitize@url \@href}%
\providecommand \@href[1]{\@@startlink{#1}\@@href}%
\providecommand \@@href[1]{\endgroup#1\@@endlink}%
\providecommand \@sanitize@url [0]{\catcode `\\12\catcode `\$12\catcode
  `\&12\catcode `\#12\catcode `\^12\catcode `\_12\catcode `\%12\relax}%
\providecommand \@@startlink[1]{}%
\providecommand \@@endlink[0]{}%
\providecommand \url  [0]{\begingroup\@sanitize@url \@url }%
\providecommand \@url [1]{\endgroup\@href {#1}{\urlprefix }}%
\providecommand \urlprefix  [0]{URL }%
\providecommand \Eprint [0]{\href }%
\providecommand \doibase [0]{https://doi.org/}%
\providecommand \selectlanguage [0]{\@gobble}%
\providecommand \bibinfo  [0]{\@secondoftwo}%
\providecommand \bibfield  [0]{\@secondoftwo}%
\providecommand \translation [1]{[#1]}%
\providecommand \BibitemOpen [0]{}%
\providecommand \bibitemStop [0]{}%
\providecommand \bibitemNoStop [0]{.\EOS\space}%
\providecommand \EOS [0]{\spacefactor3000\relax}%
\providecommand \BibitemShut  [1]{\csname bibitem#1\endcsname}%
\let\auto@bib@innerbib\@empty
\bibitem [{\citenamefont {Mills}\ \emph {et~al.}(2014)\citenamefont {Mills},
  \citenamefont {Kampf}, \citenamefont {Christy}, \citenamefont {Housley},
  \citenamefont {Rossman}, \citenamefont {Reynolds},\ and\ \citenamefont
  {Marty}}]{Mills2014}%
  \BibitemOpen
  \bibfield  {author} {\bibinfo {author} {\bibfnamefont {S.~J.}\ \bibnamefont
  {Mills}}, \bibinfo {author} {\bibfnamefont {A.~R.}\ \bibnamefont {Kampf}},
  \bibinfo {author} {\bibfnamefont {A.~G.}\ \bibnamefont {Christy}}, \bibinfo
  {author} {\bibfnamefont {R.~M.}\ \bibnamefont {Housley}}, \bibinfo {author}
  {\bibfnamefont {G.~R.}\ \bibnamefont {Rossman}}, \bibinfo {author}
  {\bibfnamefont {R.~E.}\ \bibnamefont {Reynolds}},\ and\ \bibinfo {author}
  {\bibfnamefont {J.}~\bibnamefont {Marty}},\ }\bibfield  {title} {\bibinfo
  {title} {Bluebellite and mojaveite, two new minerals from the central
  {M}ojave {D}esert, {C}alifornia, {USA}},\ }\href
  {https://doi.org/10.1180/minmag.2014.078.5.15} {\bibfield  {journal}
  {\bibinfo  {journal} {Mineralog. Mag.}\ }\textbf {\bibinfo {volume} {78}},\
  \bibinfo {pages} {1325–1340} (\bibinfo {year} {2014})}\BibitemShut
  {NoStop}%
\bibitem [{\citenamefont {Haraguchi}\ \emph {et~al.}(2021)\citenamefont
  {Haraguchi}, \citenamefont {Matsuo}, \citenamefont {Kindo},\ and\
  \citenamefont {Hiroi}}]{Haraguchi2021}%
  \BibitemOpen
  \bibfield  {author} {\bibinfo {author} {\bibfnamefont {Y.}~\bibnamefont
  {Haraguchi}}, \bibinfo {author} {\bibfnamefont {A.}~\bibnamefont {Matsuo}},
  \bibinfo {author} {\bibfnamefont {K.}~\bibnamefont {Kindo}},\ and\ \bibinfo
  {author} {\bibfnamefont {Z.}~\bibnamefont {Hiroi}},\ }\bibfield  {title}
  {\bibinfo {title} {Quantum antiferromagnet bluebellite comprising a
  maple-leaf lattice made of spin-1/2 {C}u$^{2+}$ ions},\ }\href
  {https://doi.org/10.1103/PhysRevB.104.174439} {\bibfield  {journal} {\bibinfo
   {journal} {Phys. Rev. B}\ }\textbf {\bibinfo {volume} {104}},\ \bibinfo
  {pages} {174439} (\bibinfo {year} {2021})}\BibitemShut {NoStop}%
\bibitem [{\citenamefont {Sachdev}\ and\ \citenamefont
  {Bhatt}(1990)}]{BOT-Sachdev}%
  \BibitemOpen
  \bibfield  {author} {\bibinfo {author} {\bibfnamefont {S.}~\bibnamefont
  {Sachdev}}\ and\ \bibinfo {author} {\bibfnamefont {R.~N.}\ \bibnamefont
  {Bhatt}},\ }\bibfield  {title} {\bibinfo {title} {Bond-operator
  representation of quantum spins: Mean-field theory of frustrated quantum
  {H}eisenberg antiferromagnets},\ }\href
  {https://doi.org/10.1103/PhysRevB.41.9323} {\bibfield  {journal} {\bibinfo
  {journal} {Phys. Rev. B}\ }\textbf {\bibinfo {volume} {41}},\ \bibinfo
  {pages} {9323} (\bibinfo {year} {1990})}\BibitemShut {NoStop}%
\bibitem [{\citenamefont {Normand}\ and\ \citenamefont
  {R\"uegg}(2011)}]{Normand2011}%
  \BibitemOpen
  \bibfield  {author} {\bibinfo {author} {\bibfnamefont {B.}~\bibnamefont
  {Normand}}\ and\ \bibinfo {author} {\bibfnamefont {C.}~\bibnamefont
  {R\"uegg}},\ }\bibfield  {title} {\bibinfo {title} {Complete bond-operator
  theory of the two-chain spin ladder},\ }\href
  {https://doi.org/10.1103/PhysRevB.83.054415} {\bibfield  {journal} {\bibinfo
  {journal} {Phys. Rev. B}\ }\textbf {\bibinfo {volume} {83}},\ \bibinfo
  {pages} {054415} (\bibinfo {year} {2011})}\BibitemShut {NoStop}%
\bibitem [{\citenamefont {Fishman}\ \emph {et~al.}(2022)\citenamefont
  {Fishman}, \citenamefont {White},\ and\ \citenamefont
  {Stoudenmire}}]{itensor}%
  \BibitemOpen
  \bibfield  {author} {\bibinfo {author} {\bibfnamefont {M.}~\bibnamefont
  {Fishman}}, \bibinfo {author} {\bibfnamefont {S.~R.}\ \bibnamefont {White}},\
  and\ \bibinfo {author} {\bibfnamefont {E.~M.}\ \bibnamefont {Stoudenmire}},\
  }\bibfield  {title} {\bibinfo {title} {{The ITensor Software Library for
  Tensor Network Calculations}},\ }\href
  {https://doi.org/10.21468/SciPostPhysCodeb.4} {\bibfield  {journal} {\bibinfo
   {journal} {SciPost Phys. Codebases}\ ,\ \bibinfo {pages} {4}} (\bibinfo
  {year} {2022})}\BibitemShut {NoStop}%
\bibitem [{\citenamefont {Lyons}\ and\ \citenamefont
  {Kaplan}(1960)}]{Lyons1960}%
  \BibitemOpen
  \bibfield  {author} {\bibinfo {author} {\bibfnamefont {D.~H.}\ \bibnamefont
  {Lyons}}\ and\ \bibinfo {author} {\bibfnamefont {T.~A.}\ \bibnamefont
  {Kaplan}},\ }\bibfield  {title} {\bibinfo {title} {{Method for determining
  ground-state spin configurations}},\ }\href
  {https://doi.org/10.1103/PhysRev.120.1580} {\bibfield  {journal} {\bibinfo
  {journal} {Phys. Rev.}\ }\textbf {\bibinfo {volume} {120}},\ \bibinfo {pages}
  {1580} (\bibinfo {year} {1960})}\BibitemShut {NoStop}%
\bibitem [{\citenamefont {Kaplan}\ and\ \citenamefont
  {Menyuk}(2007)}]{Kaplan2007}%
  \BibitemOpen
  \bibfield  {author} {\bibinfo {author} {\bibfnamefont {T.~A.}\ \bibnamefont
  {Kaplan}}\ and\ \bibinfo {author} {\bibfnamefont {N.}~\bibnamefont
  {Menyuk}},\ }\bibfield  {title} {\bibinfo {title} {{Spin ordering in
  three-dimensional crystals with strong competing exchange interactions}},\
  }\href {https://doi.org/10.1080/14786430601080229} {\bibfield  {journal}
  {\bibinfo  {journal} {Phil. Mag.}\ }\textbf {\bibinfo {volume} {87}},\
  \bibinfo {pages} {3711} (\bibinfo {year} {2007})}\BibitemShut {NoStop}%
\bibitem [{\citenamefont {Brown}\ \emph {et~al.}(2006)\citenamefont {Brown},
  \citenamefont {Fox}, \citenamefont {Maslen}, \citenamefont
  {O{\textquotesingle}Keefe},\ and\ \citenamefont {Willis}}]{Brown2006}%
  \BibitemOpen
  \bibfield  {author} {\bibinfo {author} {\bibfnamefont {P.~J.}\ \bibnamefont
  {Brown}}, \bibinfo {author} {\bibfnamefont {A.~G.}\ \bibnamefont {Fox}},
  \bibinfo {author} {\bibfnamefont {E.~N.}\ \bibnamefont {Maslen}}, \bibinfo
  {author} {\bibfnamefont {M.~A.}\ \bibnamefont {O{\textquotesingle}Keefe}},\
  and\ \bibinfo {author} {\bibfnamefont {B.~T.~M.}\ \bibnamefont {Willis}},\
  }\bibfield  {title} {\bibinfo {title} {Intensity of diffracted intensities},\
  }in\ \href {https://doi.org/10.1107/97809553602060000600} {\emph {\bibinfo
  {booktitle} {International Tables for Crystallography}}}\ (\bibinfo
  {publisher} {International Union of Crystallography},\ \bibinfo {year}
  {2006})\ pp.\ \bibinfo {pages} {554--595}\BibitemShut {NoStop}%
\end{thebibliography}%

\end{document}


\title{Supplemental Materials: Effective spin-1 breathing kagome Hamiltonian induced by the exchange hierarchy in the maple leaf mineral bluebellite}
\author{Pratyay Ghosh}
\affiliation{Institut f\"ur Theoretische Physik und Astrophysik and W\"urzburg-Dresden Cluster of Excellence ct.qmat, Universit\"at W\"urzburg,
Am Hubland Campus S\"ud, W\"urzburg 97074, Germany}

\author{Tobias M\"uller}
\affiliation{Institut f\"ur Theoretische Physik und Astrophysik and W\"urzburg-Dresden Cluster of Excellence ct.qmat, Universit\"at W\"urzburg,
Am Hubland Campus S\"ud, W\"urzburg 97074, Germany}

\author{Yasir Iqbal}
\affiliation{Department of Physics and Quantum Centers in Diamond and Emerging Materials (QuCenDiEM) group, Indian Institute of Technology Madras, Chennai 600036, India}

\author{Ronny Thomale}
\affiliation{Institut f\"ur Theoretische Physik und Astrophysik and W\"urzburg-Dresden Cluster of Excellence ct.qmat, Julius-Maximilians-Universit\"at W\"urzburg, Am Hubland Campus S\"ud, W\"urzburg 97074, Germany}
\affiliation{Department of Physics and Quantum Centers in Diamond and Emerging Materials (QuCenDiEM) group, Indian Institute of Technology Madras, Chennai 600036, India}

\author{Harald O. Jeschke}
\affiliation{Research Institute for Interdisciplinary Science, Okayama University, Okayama 700-8530, Japan}
\affiliation{Department of Physics and Quantum Centers in Diamond and Emerging Materials (QuCenDiEM) group, Indian Institute of Technology Madras, Chennai 600036, India}

\maketitle
\beginsupplement

\section{Details of electronic structure calculations and energy mapping}

The crystal structure of bluebellite {\bb} has been determined from mineral samples~\cite{Mills2014} and from synthetic polycrystals~\cite{Haraguchi2021}. In the latter case, hydrogen positions were not determined. As the magnetic measurements were performed for synthetic bluebellite, it is necessary to prepare the crystal structure for electronic structure calculations and energy mapping by adding and optimizing hydrogen positions. Furthermore, Cl, O2 and H2 all have coordinates (0,0,z), restricting their position by symmetry to an axis, and H2 needs to be placed between Cl and O2, in accordance with the Mills {\it et al.} structure~\cite{Mills2014}. At a distance $d_{\rm Cl-O2}=2.205$\,{\AA}, there is insufficient space for an O2-H2 bond and a Cl-H hydrogen bond. Therefore, Cl and O2 positions need to be optimized as well. For consistency, we optimize all O positions as well. The resulting structure is given in Table~\ref{tab:structure}. 

\begin{table}[htb]
\centering
\caption{Crystal structure of bluebellite {\bb} with DFT optimized Cl, O and H positions. The lattice parameters of space group $R3$ (No. 146) were kept fixed at experimental values $a=8.3056$\,{\AA} and $c=13.2194$\,{\AA}~\cite{Haraguchi2021}.}\label{tab:structure}
{\sffamily 
  \begin{tabular}{c|d{3.7}|d{3.7}|d{3.7}}
         atom   & $x$ & $y$ & $z$ \\\hline
   Cu1 & 0.4578  & 0.3867& 0.2901\\
   Cu2 & 0.0261 & 0.2404& 0.2747\\
   I   & 0.0000& 0.0000& 0.6050\\
   O1  &-0.244716&-0.154414&-0.120656\\
   O2  & 0.000000& 0.000000& 0.249551\\
   O3  & 0.190821& 0.445737& 0.171976\\
   O4  &-0.444508&-0.062377& 0.031739\\
   O5  & 0.412671& 0.135042& 0.022041\\
   Cl  & 0.000000& 0.000000& 0.049719\\
   H1  &-0.260301&-0.167341&-0.197392\\
   H2  & 0.000000& 0.000000& 0.172600\\
   H3  &-0.114255&-0.306030&-0.254043\\
   H4  &-0.454172&-0.185148& 0.109447
\end{tabular}
}
\end{table}

\begin{figure}
    \centering
    \includegraphics[width=\columnwidth]{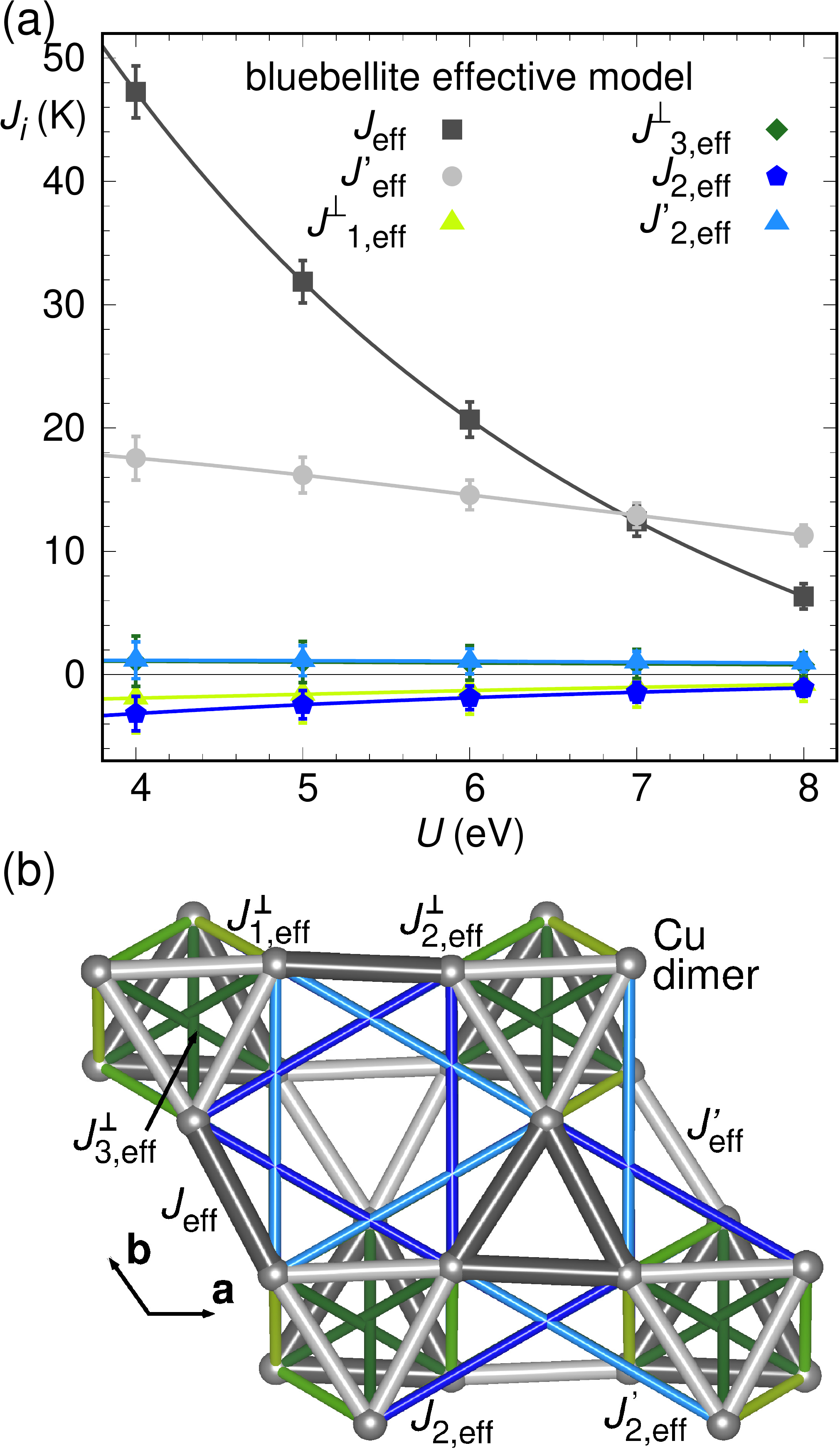}
    \caption{Energy mapping calculation for the effective spin-1 kagome lattice Hamiltonian.}
    \label{fig:S1}
\end{figure}

As explained in the main text, the large ferromagnetic $J_1$ bonds of bluebellite lead to an effective spin-1 breathing kagome lattice behavior. In order to determine the exchange interactions of this effective model, we performed energy mapping in a $3\times 1\times 1$ supercell with 18 Cu sites where we constrained moments adjacent to the $J_1$ bond to be parallel. In this way, we determine the effective exchange interaction for the spin-1 breathing kagome lattice. As shown in Fig.~\ref{fig:S1}, we perform the energy mapping for five different $U$ values. We use the same $U$ value that is relevant for the original maple leaf lattice (see main text, Fig. 1) to determine the effective Hamiltonian parameters. The result is $J_{\rm eff}= 49(2)$\,K, $J'_{\rm eff}=  18(2)$\,K, $J^\perp_{\rm 1,eff}= -2(3)$\,K, $J^\perp_{\rm 3,eff}=  1(2)$\,K, $J_{\rm 2,eff}= -3(2)$\,K, $J'_{\rm 2,eff}= 1(2)$\,K. Due to the error bars, not so much information about the subleading couplings can be obtained. It is clear that the effective kagome lattice has only very small interlayer couplings $J^\perp_{i,\rm eff}$. The inplane second neighbor couplings on average appear to be slightly ferromagnetic.

\section{Details of bond-operator mean-field theory}
\begin{figure}[b]
    \centering
    \includegraphics[width=\columnwidth]{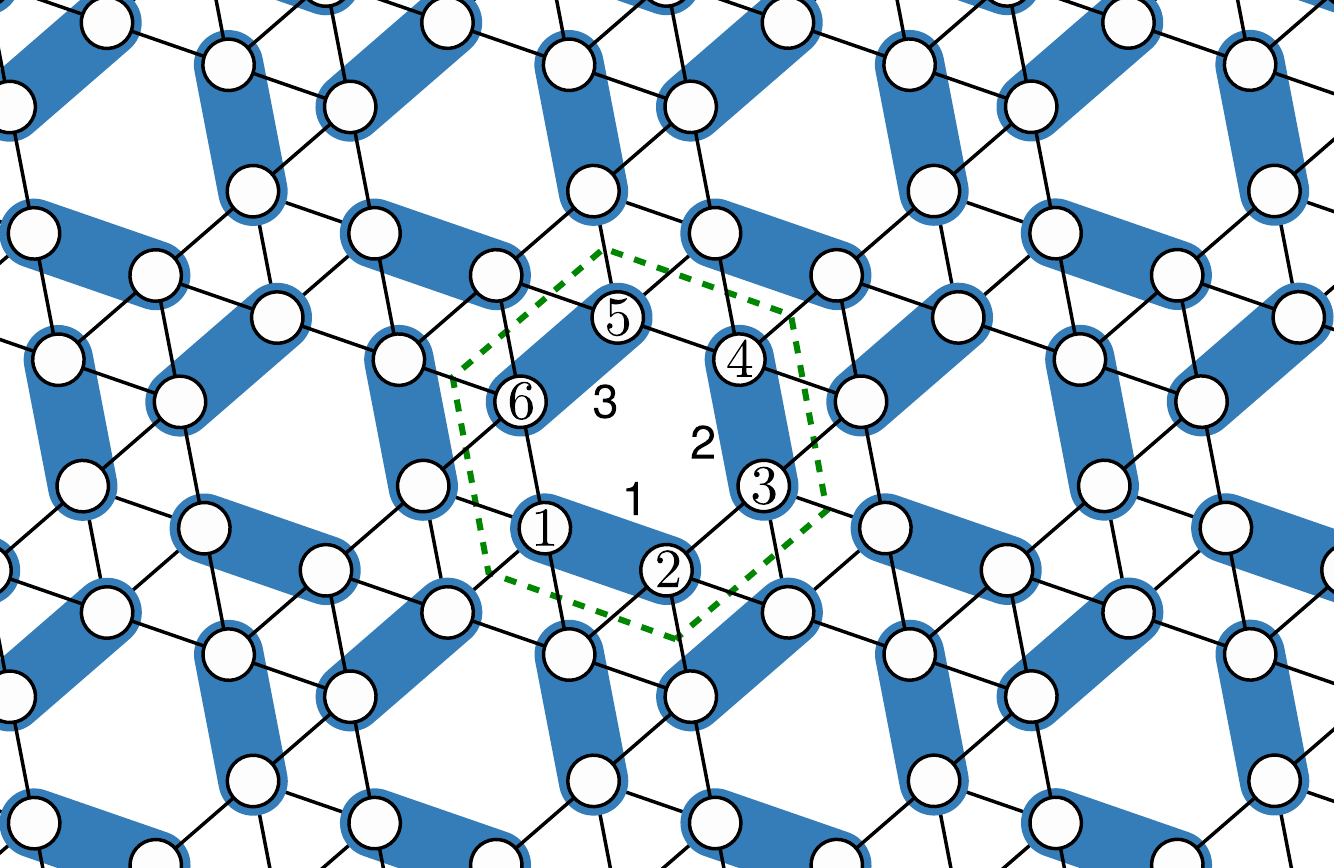}
    \caption{The singlet product state used in our bond-operator mean-field calculation. The singlets are forming on the $J_2$ bonds. We use the dashed green hexagon as our unit cell, which contains three symmetry related dimers. The indexing of the bonds and the sites in the unit cell are also marked.}
    \label{fig:S2}
\end{figure}
Based on our findings from the DMRG calculations of the Hamiltonian for bluebellite, we now develop an effective low-energy bosonic theory for various reasons. First we want to obtain an better understanding of the system at the thermodynamic limit. Secondly, such an approximate theory allows us to calculate static and dynamic spin structure factor of the system very easily. Additionally, we can gain some insight of the thermodynamic properties of the system as well.    

Upon carefully observing the NN spin-spin correlations obtained from DMRG (See Fig. 1 of main text), we propose that a minimal low-energy physics of the system can be well described by assuming that the ground state of the system is a dimerized singlet with strong singlet weights on the $J_2$ bonds. To understand the effective low-energy physics of the system, we start with a $J_2-J_3$ hexagon as our unit cell, and only consider the three $J_2$ bonds as our elementary block, $\begin{gathered}\includegraphics[scale=0.14]{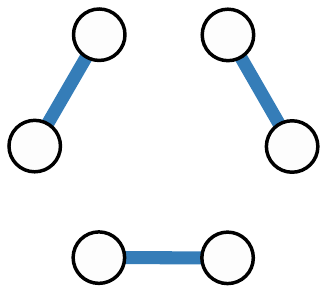}\end{gathered}$, described by the Hamiltonian
\bea\label{eq-hamil-tri}
\mathcal{H}_{\begin{gathered}\includegraphics[scale=0.14]{./t1.pdf}\end{gathered}}&=&J_2\sum_{b=1,2,3}\left(\vec{S}_{2b-1}\cdot\vec{S}_{2b}\right)\\
&=&J_2\left(\vec{S}_1\cdot\vec{S}_2+\vec{S}_3\cdot\vec{S}_4+\vec{S}_5\cdot\vec{S}_6\right),
\eea
where $b=1,2,3$ is the bond index (see Fig.\ref{fig:S2}). The ground state of this Hamiltonian is a product state of singlets forming on the $1\text{-}2$, $3\text{-}4$, and $5\text{-}6$ spin pairs, which allows us to use the bond-operator formalism~\cite{BOT-Sachdev} to represent the spin operators as
\begin{subequations}
\be
S_{2b-1}^\alpha=-\frac{1}{2}\left(\hats_{b}\hat{t}_{b}^{\alpha\dagger}+\hats_{b}^{\dagger}\hat{t}_{b}^\alpha\right)\\
\ee 
\be
S_{2b}^\alpha=\frac{1}{2}\left(\hats_{b}\hat{t}_{b}^{\alpha\dagger}+\hats_{b}^{\dagger}\hat{t}_{b}^\alpha\right).
\ee
\end{subequations}
In writing the above representation, one makes use of the basis of the singlet, $|s_b\rangle$, and three triplets, $|t^{\pm 1,0}_b\rangle$, defined on the bond $b$. On a Fock space with vaccum, $
|\varnothing\rangle_b$, the singlet and the triplet operators are defined as  
\begin{subequations}
\be
|s_b\rangle=\hats_b^\dagger|\varnothing\rangle_b
\ee 
\be
|t^m_b\rangle=\hatt^{m\dagger}_{b}|\varnothing\rangle_b,
\ee
\end{subequations}
with $\hats_b$ and $\hatt^m_{b}$ being bosonic operators. A boson number constraint 
\be\label{eq-cons}
\hat{s}_b^\dagger \hat{s}_b+\sum_{m=-1,0,1}\hatt^{m\dagger}_b\hatt^m_{b}=1
\ee
must also be satisfied on every bond. In terms of the singlet and he triplet operators defined above $\mathcal{H}_{\begin{gathered}\includegraphics[scale=0.14]{./t1.pdf}\end{gathered}}$ reads as,
\be 
\mathcal{H}_{\begin{gathered}\includegraphics[scale=0.14]{./t1.pdf}\end{gathered}}=-\frac{3}{4}J_2\sum_{b}\hats_b^\dagger \hats_b+\frac{1}{4}J_2\sum_{b}\sum_{\alpha=x,y,z} \hatt^{\alpha\dagger}_b\hatt^\alpha_{b}
\ee
where 
\bes
\be
\hatt^{x\dagger}=\frac{1}{\sqrt{2}}\left(\hat{t}^{-1\dagger}_b-\hat{t}^{1\dagger}_b\right)
\ee 
\be
\hatt^{y\dagger}=\frac{i}{\sqrt{2}}\left(\hat{t}^{-1\dagger}_b+\hat{t}^{1\dagger}_b\right)
\ee
\be
\hatt^{z\dagger}=\hat{t}^{0\dagger}_b.
\ee
\ees
Next, we rewrite our full Hamiltonian and recast it in terms of the ``coordinate" operator
\be 
\hat{Q}_{b}^{\alpha}=\frac{1}{\sqrt{2}}\left(\hat{t}_{i}^{\alpha\dagger}+\hat{t}_{i}^\alpha\right)
\ee
and its conjugate momentum operator
\be 
\hat{P}_{b}^{\alpha}=\frac{i}{\sqrt{2}}\left(\hat{t}_{i}^{\alpha\dagger}-\hat{t}_{i}^\alpha\right).
\ee
Thus, the final form of the full Hamiltonian, $\mathcal{H}$, on $N_{\rm uc}$ unit-cells reads as,
\be
\mathcal{H}\approx \mathcal{H}_{\text{MF}}= e_0 N_{\rm uc}+\frac{1}{2}\sum_{\bf{k}}\sum_{\alpha}\left[\lambda \hat{\bf{P}}_{\bf{k}}^{\alpha\dagger}\hat{\bf{P}}_{\bf{k}}^{\alpha}+\hat{\bf{Q}}_{\bf{k}}^{\alpha\dagger}\mathcal{V}_{\bf{k}}^\alpha\hat{\bf{Q}}_{\bf{k}}^{\alpha}\right].
\ee
Here, $e_0=-3J_2\overline{s}^2+\frac{3}{4}J_2+3\lambda\overline{s}^2-\frac{15}{2}\lambda$, with $\overline{s}$ being the mean singlet amplitude on all the $J_2$ bonds. $\lambda$ is the Lagrange multiplier used to satisfy the boson number constraint in \eqref{eq-cons} on average. 
\bea
\hat{\bf{P}}_{\bf{k}}^{\alpha\dagger}=\left[\hat{P}_{1\bf{k}}^{\alpha\dagger}\text{ }\hat{P}_{2\bf{k}}^{\alpha\dagger}\text{ }\hat{P}_{3\bf{k}}^{\alpha\dagger}\right]\\
\hat{\bf{Q}}_{\bf{k}}^{\alpha\dagger}=\left[\hat{Q}_{1\bf{k}}^{\alpha\dagger}\text{ }\hat{Q}_{2\bf{k}}^{\alpha\dagger}\text{ }\hat{Q}_{3\bf{k}}^{\alpha\dagger}\right]
\eea
and 
\be
\mathcal{V}_{\bf{k}}^\alpha=\begin{bmatrix}
\lambda & \eta_{12} & \eta_{31}^\ast \\
\eta_{12}^\ast & \lambda & \eta_{23} \\
\eta_{31} & \eta_{23}^\ast & \lambda 
\end{bmatrix}  
\ee 
with
\bes
\be 
\eta_{12}=\frac{\overline{s}^2}{2}\left[-J_3+(J_5-J_1)e^{i\bf{k}\cdot\bf{a}_2}+J_4 e^{i\bf{k}\cdot\bf{a}_1}\right]
\ee 
\be 
\eta_{23}=\frac{\overline{s}^2}{2}\left[-J_3+(J_5-J_1)e^{-i\bf{k}\cdot\bf{a}_1}+J_4 e^{-i\bf{k}\cdot(\bf{a}_1-\bf{a}_2})\right]
\ee 
\be 
\eta_{31}=\frac{\overline{s}^2}{2}\left[-J_3+(J_5-J_1)e^{i\bf{k}\cdot(\bf{a}_1-\bf{a}_2)}+J_4 e^{-i\bf{k}\cdot\bf{a}_2}\right].
\ee 
\ees 
Moreover, $\hat{P}^{\alpha\dagger}_{b\bf{k}}$'s and $\hat{Q}^{\alpha\dagger}_{b\bf{k}}$'s are the Fourier components of $\hat{P}^{\alpha\dagger}_{b}(\bf{r})$'s and $\hat{Q}^{\alpha\dagger}_{b}(\bf{r})$'s, respectively, i.e. $\hat{P}^{\alpha\dagger}_{b\bf{k}}=1/\sqrt{N_{\rm uc}}\sum_{\bf{k}}e^{i\mathbf{k}\cdot{\mathbf{r}}}\hat{P}^{\alpha\dagger}_{b}(\bf{r})$ and $\hat{Q}^{\alpha\dagger}_{b\bf{k}}=1/\sqrt{N_{\rm uc}}\sum_{\bf{k}}e^{i\mathbf{k}\cdot{\mathbf{r}}}\hat{Q}^{\alpha\dagger}_{b}(\bf{r})$.

$\mathcal{H}_{\text{MF}}$ now is a problem of three coupled differential equations, which one diagonalizes to obtain
\be
\mathcal{H}_{\text{MF}}=e_0 N_{\rm uc}+\sum_{m}\sum_{\bf{k}}\sum_{\alpha}\omega_{\mathbf{k},m}^{\alpha}\left(\gamma_{\mathbf{k},m}^{\alpha\dagger}\gamma_{\mathbf{k},m}^{\alpha}+\frac{1}{2}\right)
\ee 
where $\gamma_{\mathbf{k},m}^{\alpha}$ are renormalized triplon operators, and 
\be
\omega_{\mathbf{k},m}^{\alpha}=\sqrt{\lambda\left(\lambda-\frac{1}{2}\overline{s}^2\xi_{\mathbf{k},m}^{\alpha}\right)}
\ee
with
\be 
\xi_{\mathbf{k},m}^{\alpha}=2\sqrt{-\frac{p_\mathbf{k}}{3}}\cos\left[\frac{1}{3}\cos^{-1}\left(\frac{3q_\mathbf{k}}{2p_\mathbf{k}}\sqrt{-\frac{3}{p_\mathbf{k}}}-\frac{2\pi}{3}m\right)\right]
\ee 
and
\bea
p_\mathbf{k}&=&-\left(|\eta_{12}|^2+|\eta_{23}|^2+|\eta_{31}|^2\right)\\
q_\mathbf{k}&=&2\text{Re}\left(\eta_{12}\eta_{23}\eta_{31}\right).
\eea
The ground state of the system is given by the vaccum of the quasi-particles, $\gamma_{\mathbf{k},m}^{\alpha}$, i.e., the ground state energy per site of the system is given by,
\be
e_g=\frac{e_0}{6}+\frac{1}{12N_{\rm uc}}\sum_{m}\sum_{\bf{k}}\sum_{\alpha}\omega_{\mathbf{k},m}^{\alpha}\ee 
The unknown mean-field parameters, $\lambda$ and $\overline{s}^2$ are determined by minimizing $e_g$, which leads to the following self-consistent equations
\bes
\be
\lambda=J_2+\frac{1}{12N_{\rm uc}}\sum_{m}\sum_{\bf{k}}\sum_{\alpha}\frac{\lambda \xi_{\mathbf{k},m}^{\alpha}}{2\omega_{\mathbf{k},m}^{\alpha}}
\ee
\be 
\overline{s}^2=\frac{5}{2}-\frac{1}{12N_{\rm uc}}\sum_{m}\sum_{\bf{k}}\sum_{\alpha}\frac{4\lambda-\overline{s}^2 \xi_{\mathbf{k},m}^{\alpha}}{2\omega_{\mathbf{k},m}^{\alpha}}.
\ee
\ees

To access the finite temperature properties from the bond-operator mean-field theory we employ the methodology used by Normand \textit{et. al.}~\cite{Normand2011}. First of all, 
a thermal occupation function of hard–core bosons is impossible. This problem can be tackled by imposing the local hard-core constraint \eqref{eq-cons} at the global level to obtain effective statistics as,
\be
n(\omega_{\mathbf{k},m}^{\alpha},\beta)=\frac{\exp\left(-\beta\omega_{\mathbf{k},m}^{\alpha}\right)}{1+\sum_{m,\bf{k},\alpha}\exp\left(-\beta\omega_{\mathbf{k},m}^{\alpha}\right)},
\ee
where $\beta$ is the inverse temperature. The magnetic specific heat, hereafter, is easily derived by taking a second derivative of the free-energy with respect to temperature. The magnetic specific heat thus obtained reads as,
\begin{widetext}
\be
C_{\text{mag}}(\beta)=\sum_{m,\bf{k},\alpha}\left[\frac{\left(\beta\omega_{\mathbf{k},m}^{\alpha}\right)^{2}\exp\left(-\beta\omega_{\mathbf{k},m}^{\alpha}\right)}{1+\sum_{m,\bf{k},\alpha}\exp\left(-\beta\omega_{\mathbf{k},m}^{\alpha}\right)}-\left\{\frac{\beta\omega_{\mathbf{k},m}^{\alpha}\exp\left(-\beta\omega_{\mathbf{k},m}^{\alpha}\right)}{1+\sum_{m,\bf{k},\alpha}\exp\left(-\beta\omega_{\mathbf{k},m}^{\alpha}\right)}\right\}^{2}\right].
\ee 
\end{widetext}
The results obtained from the finite temperature calculations are depicted in Fig.~3 of the main text.

\section{Further results from DMRG}
\begin{figure*}
    \centering
    \includegraphics[width=\textwidth]{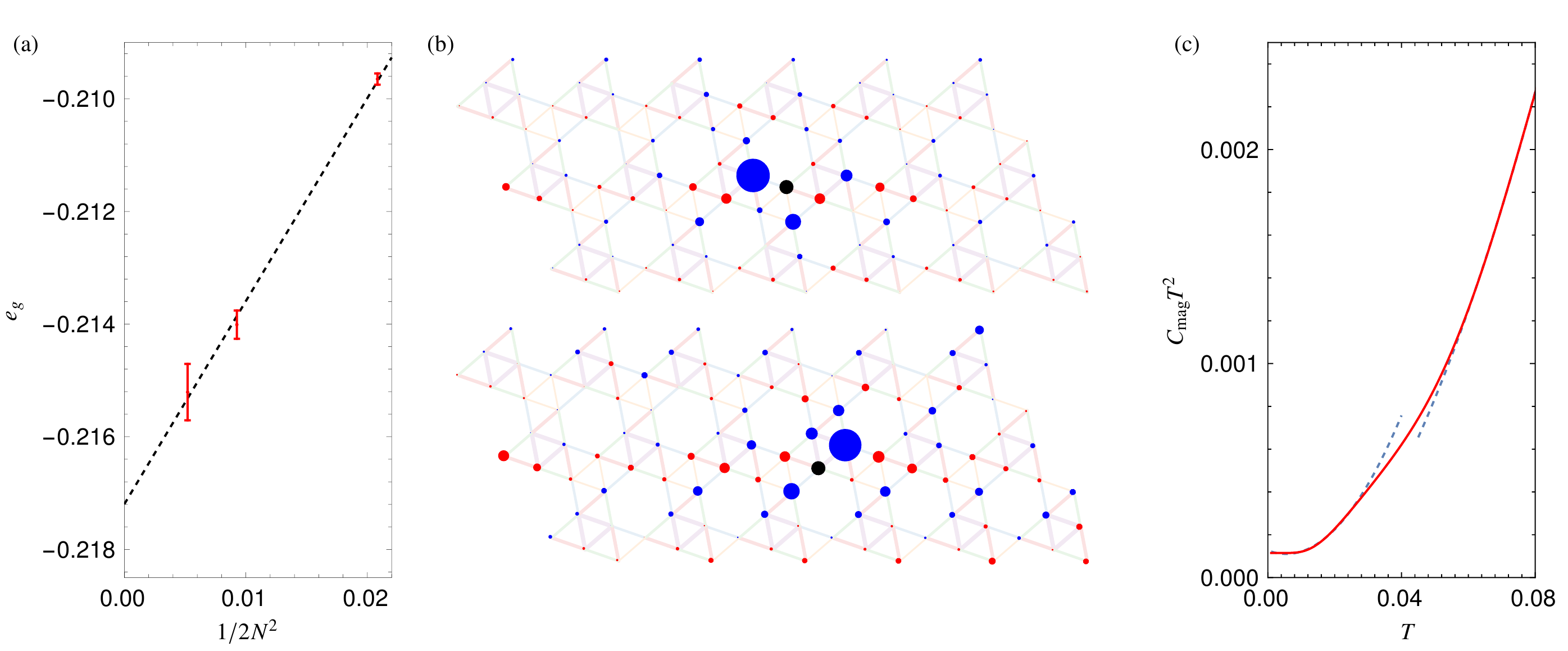}
    \caption{(a) The ground state energy per site ($e_g$) found in DMRG for system sizes $N=48, 108, 192$. The ground state energy in the thermodynamic limit is estimated to be $E_0/\widetilde{J}\approx -0.217(1)$ per site using the energy scale of $\widetilde{J}=\sqrt{J_1^2+J_2^2+J_3^2+J_4^2+J_5^2}=238\text{ K}$. (b) The spin-spin correlation $\langle\vec{S}_j\cdot\vec{S}_r\rangle$ obtained from DMRG on a 108 site maple-leaf cluster. $\vec{S}_r$ is our reference spin which is marked in black. For the top panel we set $r=63$ and for the bottom panel $r=62$. The radius of the disks indicates the strength of the correlation and the color red (blue) indicates positive (negative) correlation. In both cases we see that in the bulk the spin-spin correlation decays very quickly, signalling a spin disordered ground state. (c) The $C_{\text{mag}}T^2$ vs. $T$ calculated via DMRG on a $48$ site cluster. The calculations are done with $24$ sweeps with a maximum bond dimension of $1024$. Apart from the ground state, we also calculate $480$ exited states to calculate the finite temperature properties. The dashed line is guide for the eye. One can see an anomalous behavior occurring at $\sim 0.04\widetilde{J} =10\text{ K}$ which is also seen in the experiments.}
    \label{fig:S3}
\end{figure*}

The ground state energy of the bluebellite Hamiltonian, introduced in the main text, is calculated via DMRG using the ITENSOR~\cite{itensor} package. For the calculations, we use three different sized spin tubes with $N=2$, $3$, and $4$ unit cells along the circumference of the tube. Along the length of the tube we always take $2N$ unit cells. Thus we do our calculations on $L=48, 108, 192$ site clusters. For $L=48$ and $108$, we perform $30$ sweeps with a maximum bond dimension $2048$. For $L=192$ the number of sweeps was reduced to $24$. The ground state energy obtained via these calculations are shown in Fig.~\ref{fig:S3} (a). We perform a linear fitting of the energies to obtain a finite size scaling. The spin-spin correlations from two central sites are shown in Fig.~\ref{fig:S3} (b). A fast decay of the long range spin-spin correlations is apparent there. Next, to access the thermodynamic properties of the system, we calculate the magnetic specific heat by using,
\be
C_{\text{mag}}\propto\frac{1}{LT^2}\left[\langle\hat{H}^2\rangle-\langle\hat{H}\rangle^2\right]
\ee
where, $T$ is the temperature of the system and $\langle\hat{Q}\rangle$ is the thermal average of an observable $\hat{Q}$. For this part of the calculation we choose the $48$ site cluster. We perform $24$ sweeps with a maximum bond dimension of $512$ and calculate $480$ exited states. We estimate that for a $L$ site cluster there will be $L/2$ strong singlets forming on the $J_2$ bonds. Each such singlet would have $3$ excited triplet states. Therefore, we need more than $3L/2$ excited states to capture the finite temperature physics which will be missed by the bond-operator calculations. Thus it is an extremely costly calculation and also prone to numerical errors, and thus we only perform it on a small cluster to get an idea of the thermodynamic properties of the system which we show in Fig.~\ref{fig:S3} (c). We see an anomalous behavior occurring at $\sim 10\text{ K}$ which becomes exceedingly pronounced with increasing number of excited states considered. We believe this comes from the thermal activation of the excited states that live above the triplets. We believe the anomalous behavior seen in the experiments at $17$ K is also related to that. 

\section{Luttinger-Tisza analysis of the bluebellite}
\begin{figure}
    \centering
    \includegraphics[width=0.8\columnwidth]{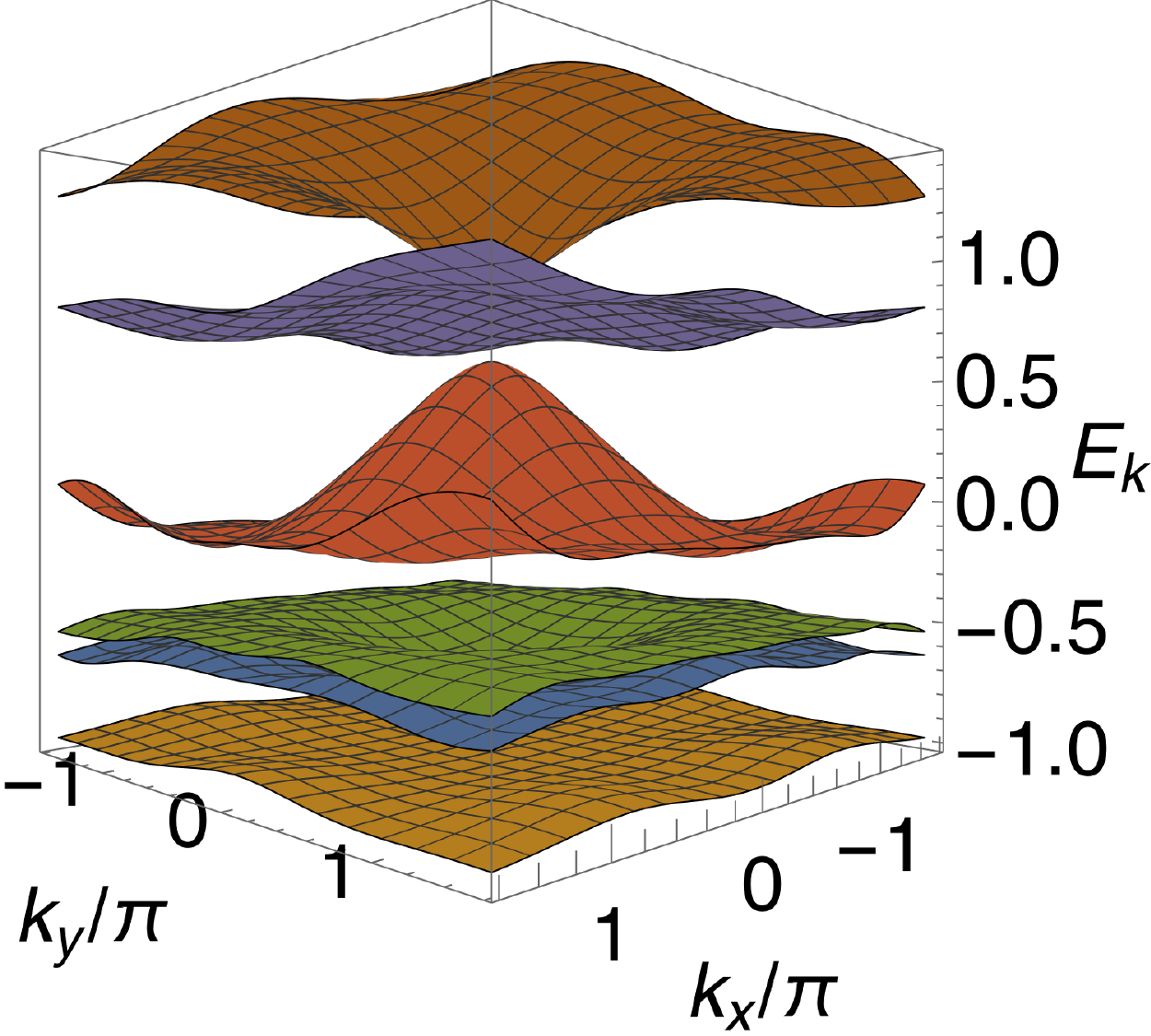}
    \caption{Luttinger-Tisza band-structure of the full bluebellite model Hamiltonian. The lowest band has a width of of 4\% of $\Tilde{J}$ with soft minima along the $\Gamma$--$M$ direction within an almost flat area around the origin in reciprocal space.}
    \label{fig:S5}
\end{figure}

The full model Hamiltonian of bluebellite is not amenable to a solution in the classical, i.e. $S \to \infty$ limit, where spin operators are replaced by vector spins. Therefore, we resort to the Luttinger-Tisza approximation~\cite{Lyons1960,Kaplan2007}, where the normalization of the spins is enforced only on average, allowing for analytical solution of the model. The corresponding band-structure in Fig.~\ref{fig:S5} shows a flat lowest band with a band-with of 4\% of $\Tilde{J}$, and an almost degenerate area around the $\Gamma$ point, featuring soft minima along the $\Gamma$--$M$ direction. While in the classical limit, even in the Luttinger-Tisza approximation, this energy landscape implies an ordered ground state, quantum fluctuations can access these low-lying states by allowing for variations in the spin expectation values. Therefore, the flatness of the lowest Luttinger-Tisza band indicates the probability for the quantum model to avoid long-range order, as we find for bluebellite.

\section{Further information for static and dynamical structure factors}
\subsection{Maple-leaf structure factor as a basis expanded kagome structure factor}
\begin{figure}
    \centering
    \includegraphics[width=0.8\columnwidth]{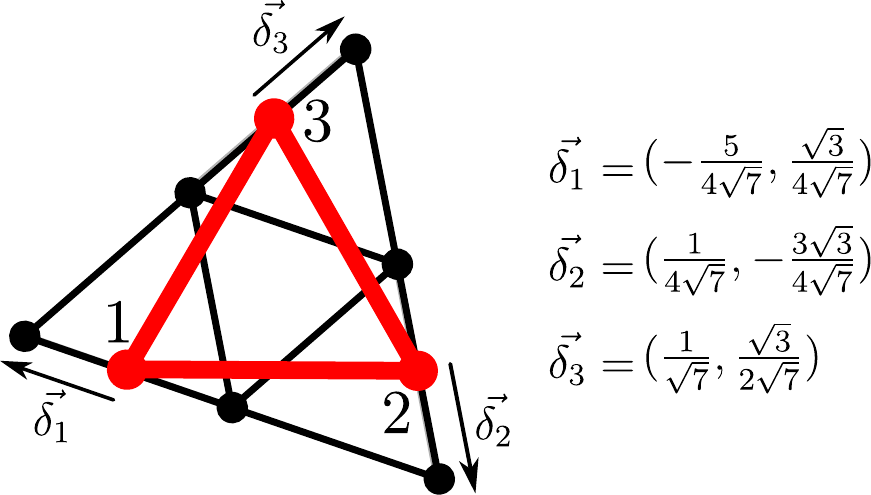}
    \caption{Basis expansion of a kagome unit-cell that produces the maple-leaf unit-cell.}
    \label{fig:S4}
\end{figure}

In the main text we have mentioned that the bluebellite Hamiltonian behaves as an effective spin-$1$ kagome system with a breathing anisotropy. Here, we demonstrate under such an assumption how one can calculate the static structure factor for the maple-leaf system from the same for a kagome system. The static structure factor for a non-Bravais system like kagome is given by, 
\be
S(\mathbf{q})=\frac{1}{N}\sum_{ij}\sum_{k^{}k^{\prime}}e^{i\mathbf{q}\cdot(\vec{R}_i-\vec{R}_j)}e^{i\mathbf{q}\cdot(\vec{b}_{k^{}}-\vec{b}_{k^{\prime}})}\langle\vec{S}_{k^{\prime}}(\vec{R}_j)\cdot\vec{S}_{k^{}}(\vec{R}_i)\rangle,
\ee 
where $\vec{R}_i$ are the positions of the unit-cells, and $\vec{b}_k$ are the basis vectors. Next, to go to the maple-leaf unit cell from the kagome unit-cell one needs to expand each basis site of kagome into two sites. For the $k$-th basis site of kagome, this is achieved by creating two sites at $\vec{b}_k\pm\vec{\delta}_{k}$ (the $\vec{\delta}_{k}$'s are given in Fig.~\ref{fig:S4}). Thereafter, the static structure factor for maple-leaf lattice is derived from the spin-spin correlations calculated for the effective kagome system reads as,
\begin{widetext}
\be
S(\mathbf{q})=\frac{1}{N}\sum_{ij}\sum_{k^{}k^{\prime}}e^{i\mathbf{q}\cdot(\vec{R}_i-\vec{R}_j)}e^{i\mathbf{q}\cdot(\vec{b}_{k^{}}-\vec{b}_{k^{\prime}})}e^{i\mathbf{q}\cdot(\pm\vec{\delta}_{k^{}}\mp\vec{\delta}_{k^{\prime}})}\langle\vec{S}_{k^{\prime}}(\vec{R}_j)\cdot\vec{S}_{k^{}}(\vec{R}_i)\rangle.
\ee 
\end{widetext}
This has been used to produce Fig.~2\,(c) in the main text, for which we start with an effective breathing kagome system with spins replaced by classical vectors. For this system the ground state is known to be degenerate with both $\mathbf{q}=0$ and $\sqrt{3}\times\sqrt{3}$ coplanar orders. For a $\sqrt{3}\times\sqrt{3}$ order, the high intensity occurs at the $K$ points of the extended Brillouin zone (BZ), while for $\mathbf{q}=0$ order, it occurs at the $M$ points of the extended BZ.

\subsection{Extended Brillioun zone} 
The form factor, $f_\mathbf{G}$, for a maple-leaf system is given by
\be
f_\mathbf{G}\propto\sum_{j}\exp(i\mathbf{G}\cdot\mathbf{v}_{j})
\ee
where $\mathbf{v}_j$ are the six basis vectors of the lattice, $G=\nu_1 \mathbf{b}_1+\nu_2 \mathbf{b}_2$ is a site on the reciprocal lattice defined by the reciprocal vectors $\mathbf{b}_1$ and $\mathbf{b}_2$ ($\nu_1$ and $\nu_2$ are integers). The maxima of $|f_\mathbf{G}|^2$ occur for all combinations of $\nu_1$ and $\nu_2$ such that $\mod(\nu_1+2\nu_2,7)=0$, i.e., by traversing along $\mathbf{b}_1$ or $\mathbf{b}_2$ the structure factor is only periodic in seven reciprocal lattice spacings. Therefore, the extended BZ of the structure factor is seven times bigger than the actual BZ of the lattice (see Fig.~2\,(e) in the main text). 

\subsection{Magnetic form factor for Cu$^{2+}$} 
In Fig. 3 (a) of the main text, we show the magnetic form factor $|F(Q)|^2$ modulated $S(Q)$, where we use 
\bea
F(Q)=&&0.0232e^{-34.969\left(\frac{Q}{4\pi}\right)^2}+0.4023e^{-11.564\left(\frac{Q}{4\pi}\right)^2}\nonumber\\
&+&0.5882e^{-3.843\left(\frac{Q}{4\pi}\right)^2}-0.0137
\eea
for Cu$^{2+}$ ions~\cite{Brown2006}. 

\bibliography{bluebellite}